%% file: Paper_Bloch_v4.tex
\documentclass[english,aps,
pra,
superscriptaddress,
showpacs,
showkeys,
notitlepage,
twocolumn, 
numbers,
longbibliography,
nofootinbib]{revtex4-1}
\usepackage[T1]{fontenc}
\usepackage[latin9]{inputenc}
\usepackage{amsmath}
\usepackage{amssymb}
\usepackage{graphicx}
\usepackage{hyperref}
\usepackage{graphicx, xcolor}

\definecolor{darkblue}{rgb}{0,0,0.93} 
\definecolor{darkred}{rgb}{0.8,0,0} 

\hypersetup{
colorlinks=true, 
linkcolor=darkblue, 
citecolor=darkred, 
filecolor=darkblue, 
urlcolor=darkblue
}

\makeatletter
\usepackage{babel}

\makeatother

\usepackage{babel}
\begin{document}

\global\long\def\abs#1{\left|#1\right|}%
 
\global\long\def\ket#1{\left|#1\right\rangle }%
 
\global\long\def\bra#1{\left\langle #1\right|}%
 
\global\long\def\half{\frac{1}{2}}%
 
\global\long\def\partder#1#2{\frac{\partial#1}{\partial#2}}%
 
\global\long\def\comm#1#2{\left[#1,#2\right]}%
 
\global\long\def\vp{\vec{p}}%
 
\global\long\def\vpp{\vec{p}'}%
 
\global\long\def\dt#1{\delta^{(3)}(#1)}%
 
\global\long\def\Tr#1{\textrm{Tr}\left\{  #1\right\}  }%
 
\global\long\def\Real#1{\mathrm{Re}\left\{  #1 \right\}  }%
 
\global\long\def\braket#1{\langle#1\rangle}%
 
\global\long\def\escp#1#2{\left\langle #1|#2\right\rangle }%
 
\global\long\def\elmma#1#2#3{\langle#1\mid#2\mid#3\rangle}%
 
\global\long\def\ketbra#1#2{|#1\rangle\langle#2|}%

\author{Pablo Arnault}
\email{pablo.arnault@ific.uv.es}
\affiliation{Departamento de F{í}sica Te{ó}rica and IFIC, Universidad de Valencia-CSIC,Dr. Moliner
50, 46100-Burjassot, Spain}

\title{Quantum walks in weak electric fields and Bloch oscillations}
\author{Benjamin Pepper}
\affiliation{Departamento de F{í}sica Te{ó}rica and IFIC, Universidad de Valencia-CSIC,Dr. Moliner
50, 46100-Burjassot, Spain}
\affiliation{Department of Physics, Imperial College London, Blackett Laboratory, Prince Consort Road, London SW7 2AZ, U.K.}

\author{A. Pérez}
\email{armando.perez@uv.es}
\affiliation{Departamento de F{í}sica Te{ó}rica and IFIC, Universidad de Valencia-CSIC,Dr. Moliner
50, 46100-Burjassot, Spain}

\begin{abstract}
Bloch oscillations appear when an electric field is superimposed on
a quantum particle that evolves on a lattice with a tight-binding
Hamiltonian (TBH), i.e., evolves via what we will call an \emph{electric
TBH}; this phenomenon will be referred to as \emph{TBH Bloch oscillations}.
A similar phenomenon is known to show up in so-called \emph{electric
discrete-time quantum walks (DQWs)} \cite{ced13}; this phenomenon
will be referred to as \emph{DQW Bloch oscillations}. This similarity
is particularly salient when the electric field of the DQW is weak.
For a wide, i.e., spatially extended initial condition, one numerically
observes semi-classical oscillations, i.e., oscillations of a localized
particle, both for the electric TBH and the electric DQW. More precisely:
The numerical simulations strongly suggest that the semi-classical
DQW Bloch oscillations correspond to two counter-propagating semi-classical
TBH Bloch oscillations. In this work it is shown that, under certain
assumptions, the solution of the electric DQW for a weak electric
field and a wide initial condition is well approximated by the superposition
of two continuous-time expressions, which are counter-propagating
solutions of an electric TBH whose hopping amplitude is the cosine
of the \emph{arbitrary} coin-operator mixing angle. In contrast, if
one wishes the continuous-time approximation to hold for spatially
localized initial conditions, one needs at least the DQW to be lazy,
as suggested by numerical simulations and by the fact that this has
been proven in the case of a vanishing electric field \cite{Strauch06b}.
\end{abstract}
\maketitle


\section{Introduction} \label{sec:Intro}

Quantum walks can be understood as quantum analogues of classical random walks. 
In the present work, we focus on their discrete-time version, namely, discrete-time quantum walks (DQWs);
These have become very popular in recent years since they were found to have numerous applications.
One of their most relevant applications is quantum algorithmics \cite{BooleanEvalQW, ConductivityQW}, see in Ref.\ \cite{Arnault17} for a compact historical review with references to key works in this field.
Their other most relevant application, to which the present work belongs to, is the quantum simulation of physical equations and phenomena.
Indeed, DQWs can, most notably, quantum simulate the Dirac equation and, more generally, a quantum particle on a lattice in various regimes, subject to external electric and magnetic fields \cite{Bauls2006, ced13, DMD14, Bru2016, mesch13a, AD15, AD16, Yalcinkaya2015}, and/or to an external relativistic gravitational field \cite{DMD14, AD17, AF17, Arrighi2019};
A variety of high-energy-physics phenomena and situations such as neutrino oscillations \cite{Molfetta2016} and the presence of extra dimensions \cite{Bru2016b}, have also been reproduced with DQWs.

In the present work, we focus on so-called \emph{electric} DQWs on a one-dimensional (1D) lattice \cite{ced13, DMD14, Arnault17}.
These DQWs are called ``electric'' because they arise, for example and as in the present work, by implementing, at each time step, an extra, position-dependent phase, which can be interpreted, via various related aspects \cite{ced13, DMD14, AD16, Arnault17}, as an external electric potential generating an electric field.
The introduction of this phase is not necessarily due to the action of a true external electric field -- which is why one speaks of ``artificial'', or ``synthetic'' gauge field --, but it produces effects either identical or similar to the latter, depending on the considered regime.
One can show that such electric, and, in higher dimensions, electromagnetic DQWs, are invariant under a certain gauge transformation on the spacetime lattice, in relation to standard lattice gauge theories;
This lattice system and gauge invariance tend, when the appropriate limit to zero in the lattice spacing and time step is taken, to the Dirac equation coupled to an external electromagnetic field, with the well-known electromagnetic gauge invariance \cite{DMD14, AD16, MMADMP2018, APAF18, Cedzich2019}.

The dynamics arising from these electric DQWs has been considered by many authors, see the previous references.
We will focus on the case of a constant and uniform external electric field, which can be implemented by choosing the above-mentioned phase to depend \emph{linearly} on the position.
One of the observed features is the appearance of oscillations having the usual Bloch period, inversely proportional to the electric field, that we will refer to as \emph{DQW Bloch oscillations} \cite{ced13, AD16, Regensburger11, Witthaut2010}.
These DQW Bloch oscillations have been proposed as a means for the direct measurement of topological invariants \cite{PhysRevLett.118.130501}.
The standard, well-known Bloch oscillations appear when an electric field is superimposed on a quantum particle that evolves on a lattice with a tight-binding Hamiltonian (TBH) \cite{Hartmann2004, Dominguez-Adame2010, Tamascelli2016, Witthaut2004};
We will speak of \emph{electric TBHs} and \emph{TBH Bloch oscillations}.
Even if DQW Bloch oscillations have already been discussed in the literature, their relationship to TBH Bloch oscillations has not been analyzed in detail.
More precisely, we wish here to investigate the case where the oscillations appear in a semi-classical manner, i.e., show up as oscillations of a \emph{localized} particle.
For electric TBHs, it is known that an initial condition which superposes only a few lattice sites will produce, not semi-classical oscillations, but so-called ``breathing modes'', and we will see that, in this situation, similar breathing modes appear also in the electric DQW when the electric field is chosen weak, i.e., small with respect to its maximum value $\pi$.
In order to obtain semi-classical oscillations in the electric TBH, one needs a combination of many sites, such as a wide Gaussian state \cite{Hartmann2004}.
The question that arises is whether the same requirement holds for the DQW:
One can readily see numerically that yes, and that these semi-classical DQW Bloch oscillations seem to correspond to the superposition of two counter-propagating semi-classical TBH Bloch oscillations.

Now, the main achievement of the present article is to provide an analytical expression supporting these numerical observations.
To the best of our knowledge, this has not been achieved before.
As we will show, if both (i) the electric field is chosen weak, and (ii) the initial condition is wide, i.e., large with respect to the lattice spacing, one can in the end approximate the probability distribution of the electric DQW with a continuous-time analytical expression.
This expression is, both in quasimomentum and in real space, the superposition of two counter-propagating solutions of an electric TBH whose hopping amplitude is the cosine of the \emph{arbitrary} coin-operator mixing angle.
The continuous-time differential equation we derive along the way under certain assumptions, can be written in terms of a certain Hamiltonian, which takes the form of the free part of the electric TBH just mentioned.
The possibility to establish such a simple connection between the electric DQW and the electric TBH, ultimately sheds light into
the nature and behavior of DQW Bloch oscillations when the electric field is weak.
The power of the result is that we do not require the free part of the electric DQW to be lazy as in Strauch's work \cite{Strauch06b}, a situation which would be directly describable by a continuous-time quantum walk (CQW) whose Hamiltonian is simply an electric TBH.
The description here is more involved, as we will see.
The price to pay for the arbitrariness of the coin-operator mixing angle in the continuous-time approximation, is that the initial condition must be wide.

The article is organized as follows.
In Sec.\ \ref{sec:1-D-electric-quantum}, we first introduce the electric DQW and some of its basic properties, and then discuss the condition under which semi-classical both TBH Bloch oscillations and DQW Bloch oscillations appear, that is, requiring the initial condition to be wide, i.e., wide with respect to the lattice spacing.
In Sec.\ \ref{sec:Continuous-time-limit}, we define a two-step dynamics from the evolution of the electric DQW, and derive, when the electric field is weak, a continuous-time differential equation from it, which can be used to approximate the DQW evolution if the initial condition is wide.
This continuous-time differential equation can be solved, which gives continuous-time formulae that for wide initial conditions are expected to be good approximations of, and can simply be compared with the exact dynamics of the DQW, obtained by numerical simulation.
Sec.\ \ref{sec:Measure-of-probability} is devoted to quantifying the agreement between the exact probability distribution and its  continuous-time approximation.
Our main conclusions are presented in Sec.\ \ref{sec:Conclusions}.
Certain enlightening or secondary calculations have been relegated to the appendices.

\section{Electric discrete-time quantum walk on the line}  \label{sec:1-D-electric-quantum}

\subsection{Definition and basic properties}

The Hilbert space of the particle walking on the 1D lattice is  $\mathcal{H}=\mathcal{H}_{\mathrm{spatial}}\otimes\mathcal{H}_{\mathrm{coin}}$, where
$\mathcal{H}_{\mathrm{spatial}}$ is the position Hilbert space, spanned by the lattice position states, $\{\ket{x_{n}=na} / n\in\mathbb{Z} \}$, with $a>0$ the lattice spacing, and where $\mathcal{H}_{\mathrm{coin}}$ is the Hilbert space of an internal state of the walker which is called \emph{coin}, or \emph{chirality} state (see below why this term is used), that \emph{must} be introduced.
Indeed, the minimum dimension for $\mathcal{H}_{\mathrm{coin}}$ is $2$ if we want the evolution operator of the walker evolving on the spacetime lattice to be local, unitary, and translationally invariant \cite{Meyer96a}.
We work with this minimum dimension for $\mathcal{H}_{\mathrm{coin}}$ and introduce a basis of the latter, $\{\ket R, \ket L\}$, where ``$R$'' and ``$L$'' correspond to ``right' and ``left'', respectively, see below why.
We will work with the following identification, $\ket R = (1,0)^{\top}$, $\ket L = (0,1)^{\top}$, where $\top$ denotes the transposition.

The state of the particle at the discrete time $j \in \mathbb{N}$ is described by the vector $\ket{\Psi_j} \in \mathcal{H}$, and is updated via
\begin{equation} \label{eq:Psievol}
\ket{\Psi_{j+1}}= W_{\phi}(\hat{x}, \hat{k}) \ket{\Psi_j} \, ,
\end{equation}
under the action of the unitary operator
\begin{equation} \label{eq:DefW}
{W}_{\phi}(\hat{x}, \hat{k}) \equiv(e^{i\hat{x}\phi}\otimes I_{\mathrm{coin}}) {W}_{0}(\hat{k}) \, , 
\end{equation}
where
\begin{equation} \label{eq:free_walk}
{W}_{0}(\hat{k}) \equiv S(\hat{k}) (\hat{I}_{\mathrm{spatial}} \otimes C) \, .
\end{equation}
We use hats for operators acting on the position Hilbert space, but not for those acting on the coin Hilbert space.
The operators $\hat{x}$ and $\hat{k}$ are, respectively, the position and quasimomentum operators.
The operator $e^{i\hat{x}\phi}$ implements, on the free\footnote{The word ``free'' is used as usual in physics, i.e., it refers to a translationally invariant dynamics.} walk operator ${W}_{0}(\hat{k})$, this special electric field $\phi \in \mathbb{R}$ that we have talked about in the introduction, Sec.\ \ref{sec:Intro}, which is upper bounded by $2\pi/a$ \cite{ced13}.
The free walk operator is the combination of (i) an operation acting solely on the coin state, implemented by the so-called \emph{coin operator} $C$, which at this stage is an arbitrary $2\times 2$ unitary matrix, and of (ii) a coin-dependent \emph{shift operator} $S(\hat{k})$, which can be written, in the coin basis, as
\begin{equation}
S(\hat{k})\equiv e^{-i\sigma^3 \hat{k}a}=\begin{bmatrix}
e^{-i\hat{k}a} & 0\\
0 & e^{i\hat{k}a}
\end{bmatrix} ,
\end{equation}
where $\sigma^3 =\mathrm{diag}(1,-1)$ is the third Pauli matrix.
Notice that $S(\hat{k})$ shifts by one lattice site to the right the up coin component of the state, and the same but to the left for the down one, these coin components being, respectively,
\begin{subequations}  \label{eq:components}
\begin{align}
\ket {\psi^R_j} &\equiv \langle R | \Psi_j \rangle \\
\ket {\psi^L_j} &\equiv \langle L | \Psi_j \rangle \, .
\end{align}
\end{subequations}
Finally, $I_{\mathrm{coin}} =\mathbf{1}_2$, the $2\times 2$ identity matrix, and $\hat{I}_{\mathrm{spatial}}$, are respectively the identity operators acting on $\mathcal{H}_{\mathrm{coin}}$  and $\mathcal{H}_{\mathrm{spatial}}$, which we will often omit from now on to lighten the writing.
In what follows, we take the lattice spacing as the length unit, $a=1$.

Let us examine the dynamics induced by ${W}_{\phi}(\hat{x},\hat{k})$ in quasimomentum space, i.e., on the quasimomentum basis of $\mathcal{H}_{\emph{\emph{spatial}}}$, namely, $\{\ket k/\,\,k\in[-\pi,\pi[\}$, satisfying
\begin{equation}
\ket k=\sum_{n}e^{ikx_{n}}\ket{x_{n}} \, .
\end{equation}
We make use of the fact that $e^{i\hat{x}\phi}$ induces translations in quasimomentum space, $e^{i\hat{x}\phi}\ket k=\ket{k+\phi}$, whereas ${W}_{0}(\hat{k})$ is diagonal in this basis, described by a $2\times2$ matrix  $W_{0}(k)$.
Acting with $\bra k$ on the left of Eq.\ (\ref{eq:Psievol}), one arrives to 
\begin{equation} \label{eq:Dynamicsk}
\tilde{\Psi}_{j+1}(k)=W_{0}(k-\phi)\tilde{\Psi}_j(k-\phi) \, ,
\end{equation} 
where we have defined the two-component wavefunction
\begin{equation}
\tilde{\Psi}_j(k) \equiv \escp k{\Psi_j} \, .
\end{equation}
Eq.\ (\ref{eq:Dynamicsk}) shows that the dynamics of the electric DQW defined in Eq.\ \eqref{eq:DefW} can be described as the composition of two effects: The displacement in quasimomentum space, followed by the action of the free walk operator.
This dynamics has been examined both in quasimomentum space and in the original position space, and shows a rich behavior, depending on whether the ratio $\phi/(2\pi)$ is a rational or an irrational number \cite{Bauls2006, ced13, Cedzich2019bis, Bru2016}.
%

\subsection{Bloch oscillations} \label{subsec:Bloch-oscillations}

\subsubsection{Breathing modes} \label{subsubsec:breathing}

\begin{figure}
\includegraphics[width=8cm]{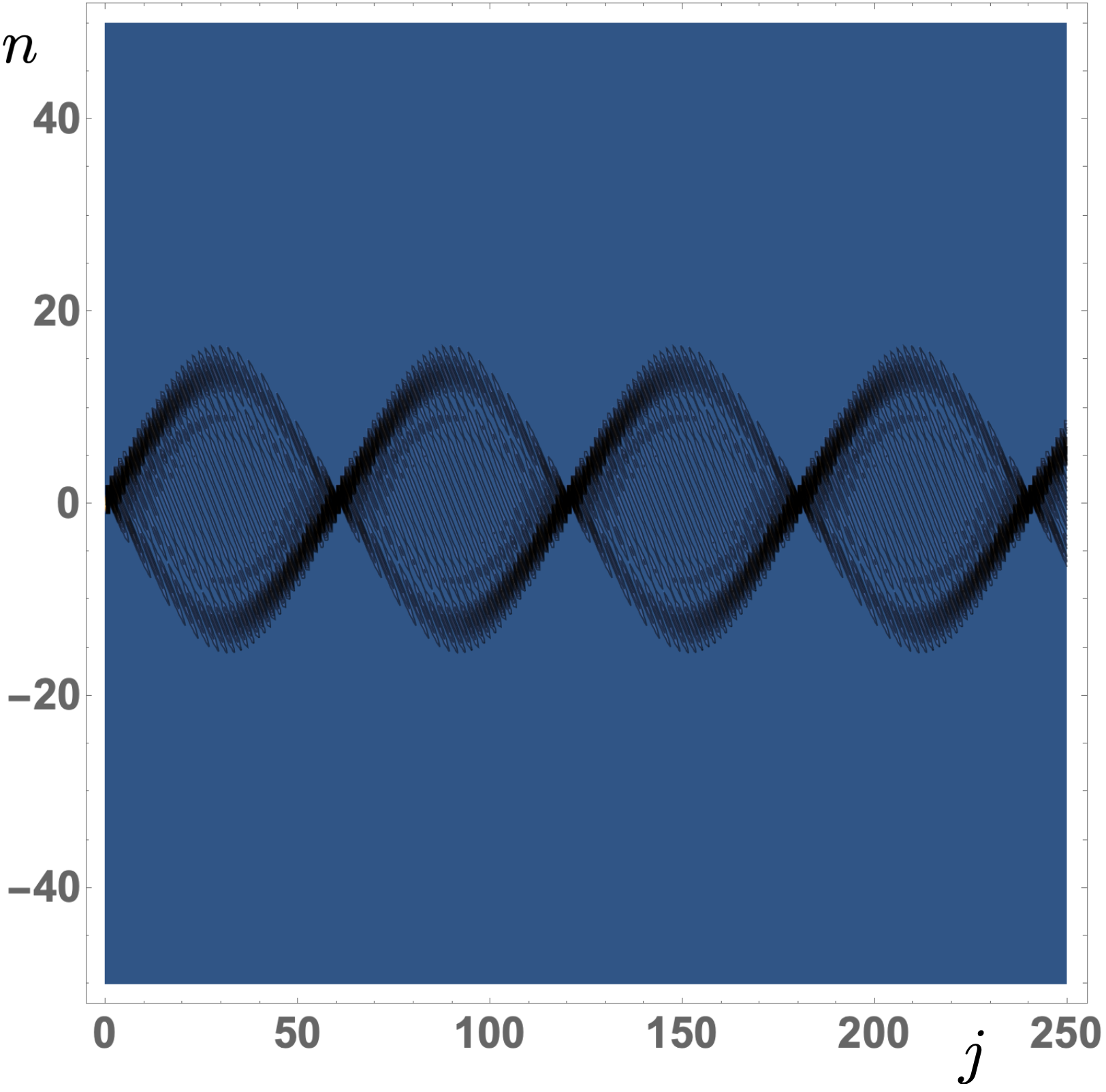}
\caption{Time evolution of the probability density $P_{j,n}$, where $n \in \mathbb{Z}$ is the lattice site and $j \in \mathbb{N}$ the time instant. The initial state of the walker is a product sate initially localized at the origin in position space, and with coin state $\frac{1}{\sqrt{2}}(1,1)^{\top}$, although almost the same plot is reproduced with the coin state $\frac{1}{\sqrt{2}}(1,-1)^{\top}$ used in next plots. The coin operator is the Hadamard one, $\theta=\pi/4$, and the electric field is $\phi=2\pi/60 \simeq 0.1$. \label{Fig:Beating_contour}}
\end{figure}
 
\begin{figure}
\begin{centering}
\includegraphics[width=8cm]{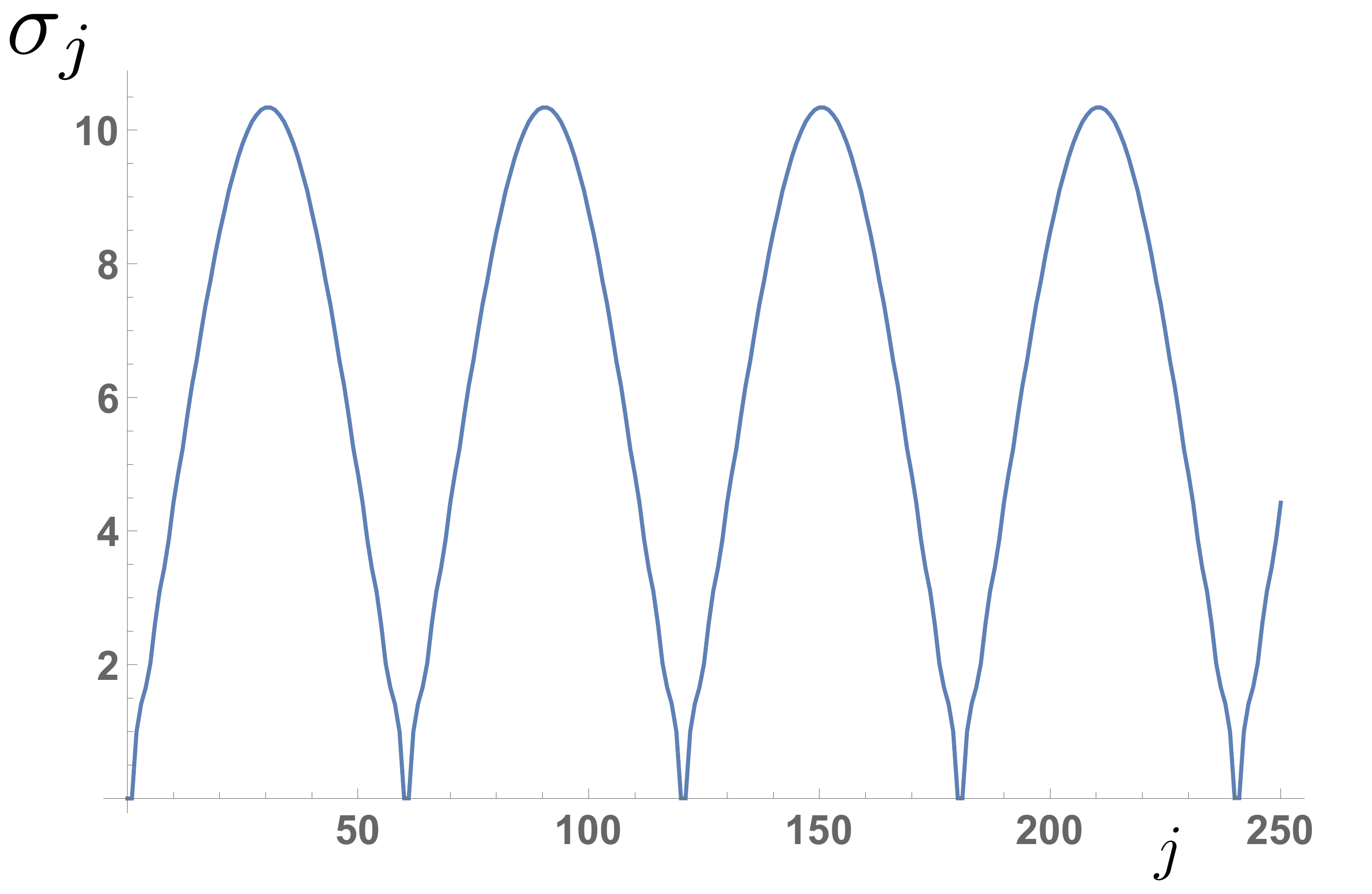}
\par\end{centering}
\caption{Time evolution of the standard deviation $\sigma_j$, where $j \in \mathbb{N}$ is the time instant, for the same conditions as in Fig. \ref{Fig:Beating_contour}.  \label{Fig:sigma_localized}}
\end{figure}

For a quantum particle with no internal degree of freedom moving on a lattice via a TBH, Bloch oscillations can be described as a consequence of the displacement in quasimomentum space due to the presence of an external constant and uniform force field, e.g., that induced by a constant and uniform electric field $E$, and they manifest as an oscillation of the probability distribution with a characteristic period $T_{\mathrm{Bloch}}(E)=2\pi/E$, the so-called \emph{Bloch period}, which is inversely proportional to the
field $E$.
For a review, see Refs.\ \cite{Hartmann2004, Dominguez-Adame2010}. 
A similar phenomenon has been observed in DQWs \cite{Cedzich2019bis, AD16}, the Bloch period being in this case $T_{\mathrm{Bloch}}(\phi)=2\pi/\phi$.
We give in the present section additional details in order to progressively motivate the main result of this work.

We choose, for the following numerical simulations, the coin operator to be
\begin{equation} \label{eq:ChoiceC}
C = \left(\begin{array}{cc}
\cos\theta & \sin\theta\\
\sin\theta & -\cos\theta
\end{array}\right) \, .
\end{equation}
Notice that the extensively used Hadamard gate corresponds to $\theta=\pi/4$. 
Numerical simulations of DQWs often use a state localized in position space as the initial condition. 
To see the effect of the electric field in this case, we run a numerical simulation of Eq.\ (\ref{eq:Psievol})  with an initial state localized at the origin.
Fig.\ \ref{Fig:Beating_contour} shows the time evolution of the probability density $P_{j,n}$, defined by
\begin{equation} \label{eq:P(n,t)}
P_{j,n}\equiv \escp{\Psi_j}{x_{n}}\escp{x_{n}}{\Psi_j} \, ,
\end{equation}
and, for the sake of completeness, we show in Fig.\ \ref{Fig:sigma_localized} the standard deviation $\sigma_j$, defined by
\begin{equation}
\sigma_j = \sqrt{\langle n^{2}\rangle_j-(\langle n\rangle_j)^{2}},
\end{equation}
where 
\begin{equation}
\langle n \rangle_{j} =\sum_{n} n P_{j,n} \, , \label{eq:n(t)}
\end{equation}
and
\begin{equation}
\langle n^{2} \rangle_{j}=\sum_{n}n^{2}P_{j,n} \, , \label{eq:n2(t)}
\end{equation}
are, respectively, the average position and average squared position at time $j$.

For the time scale we choose, we observe oscillations around the origin with a period given by the Bloch period $T_{\mathrm{Bloch}}(\phi)$, as announced.
Now, these oscillations correspond to a splitting of the initial state into two components, that tend to meet after a time $T_{\mathrm{Bloch}}(\phi)$.
A similar phenomenon also appears with electric TBHs \cite{Hartmann2004, Dominguez-Adame2010} when the initial state is localized in position space, and is referred to as ``breathing modes''.

Let us mention the differences between DQW Bloch oscillations and TBH Bloch oscillations.
First of all: In DQWs, the recombination (usually called ``revival'' \cite{ced13}) at the origin at multiples of  $T_{\mathrm{Bloch}}(\phi)$ is not perfect; This translates, in particular, into a non-vanishing standard deviation (not visible on Fig.\ \ref{Fig:sigma_localized}).
More importantly: After some time, the walker eventually moves away from the origin ballistically, i.e., asymptotically one finds $\sigma_j \propto j$.
These features, i.e., the revivals and then the ballistic excursion, actually only occur for rational values of $\phi/2\pi$ \cite{ced13}, while Anderson localization occurs for almost all irrational values \cite{Cedzich2019bis}.
For a discussion, we refer the reader to Refs.\ \cite{Bauls2006, ced13, Bru2016}.

\subsubsection{Semi-classical Bloch oscillations in position space}

\begin{figure}
\begin{centering}
\includegraphics[width=7cm]{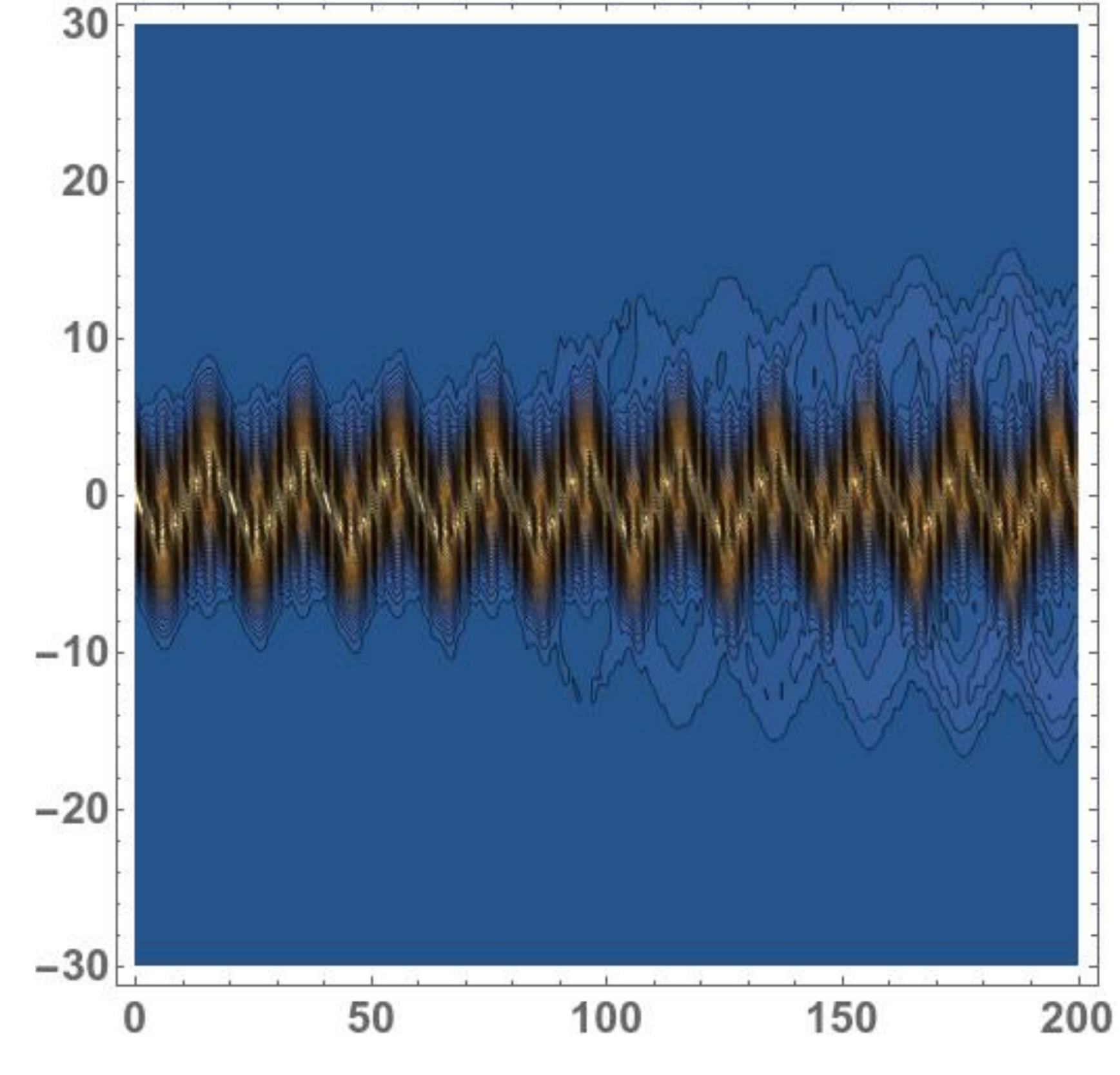}
\par\end{centering}
\begin{centering}
\includegraphics[width=7cm]{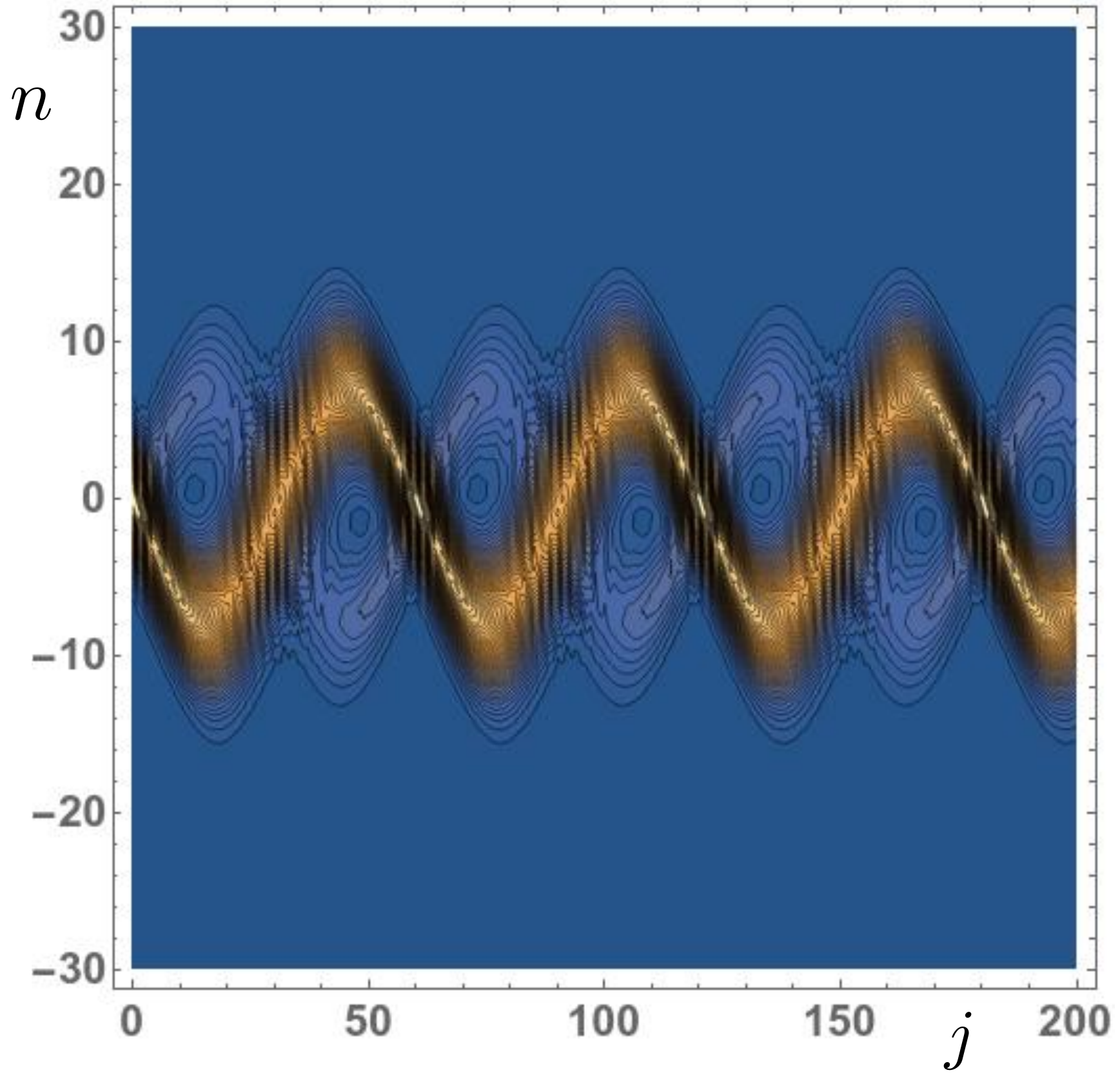}
\par\end{centering}
\caption{Contour plots corresponding to the probability density $P_{j,n}$,
as a function of the lattice site $n$ and the time instant $j$. The upper
panel is obtained with $\phi=2\pi/20$, and the lower one with $\phi=2\pi/60$.
In the simulations, $\beta=0.05$, see Eq.\ \eqref{eq:c_n}. The initial coin state is $\protect\ket s=\frac{1}{\sqrt{2}}(1,-1)$. \label{Fig:lowering_phi}}
\end{figure}

\begin{figure}
\begin{centering}
\includegraphics[width=8cm]{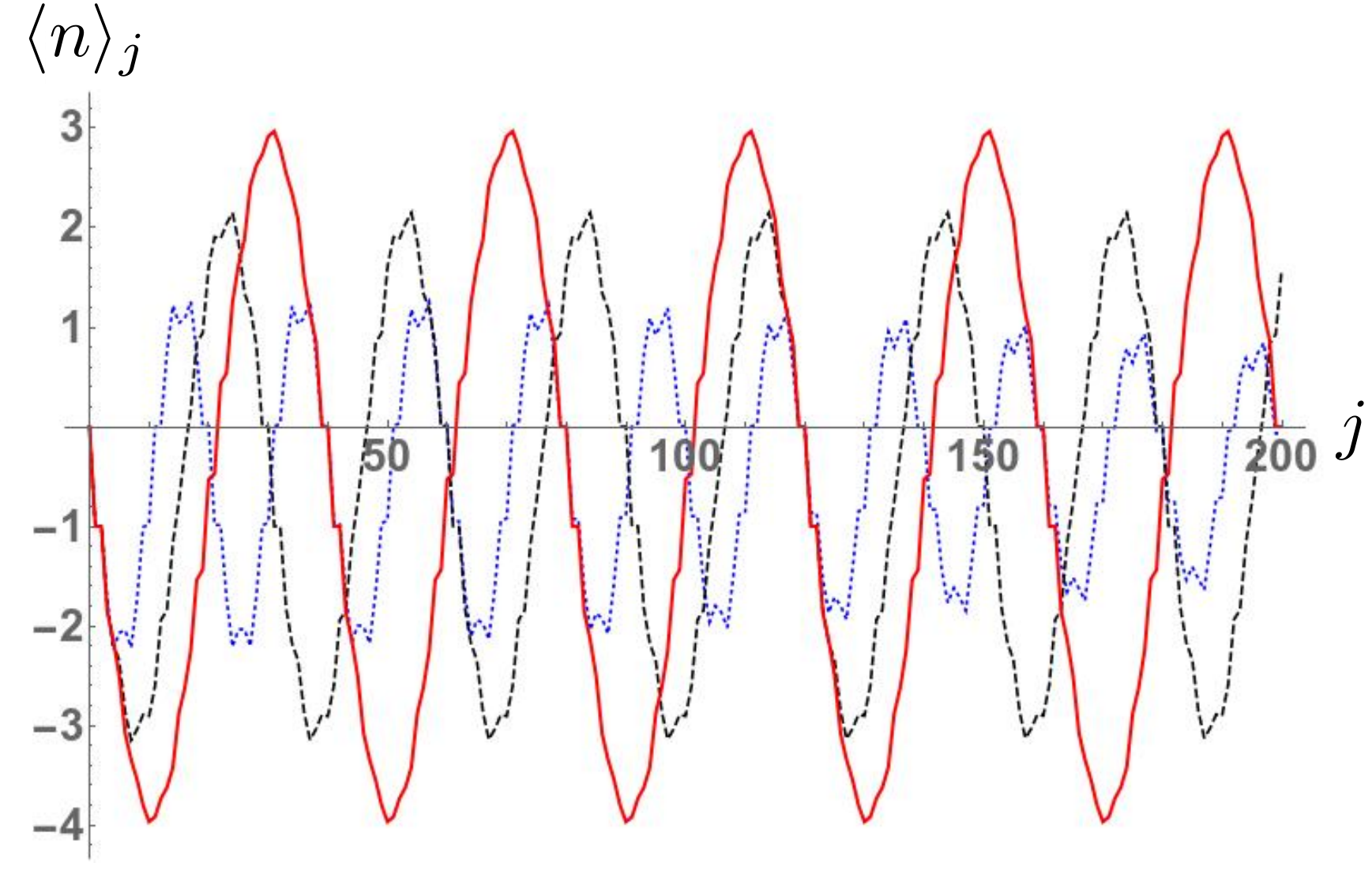} 
\par\end{centering}
\begin{centering}
\includegraphics[width=8cm]{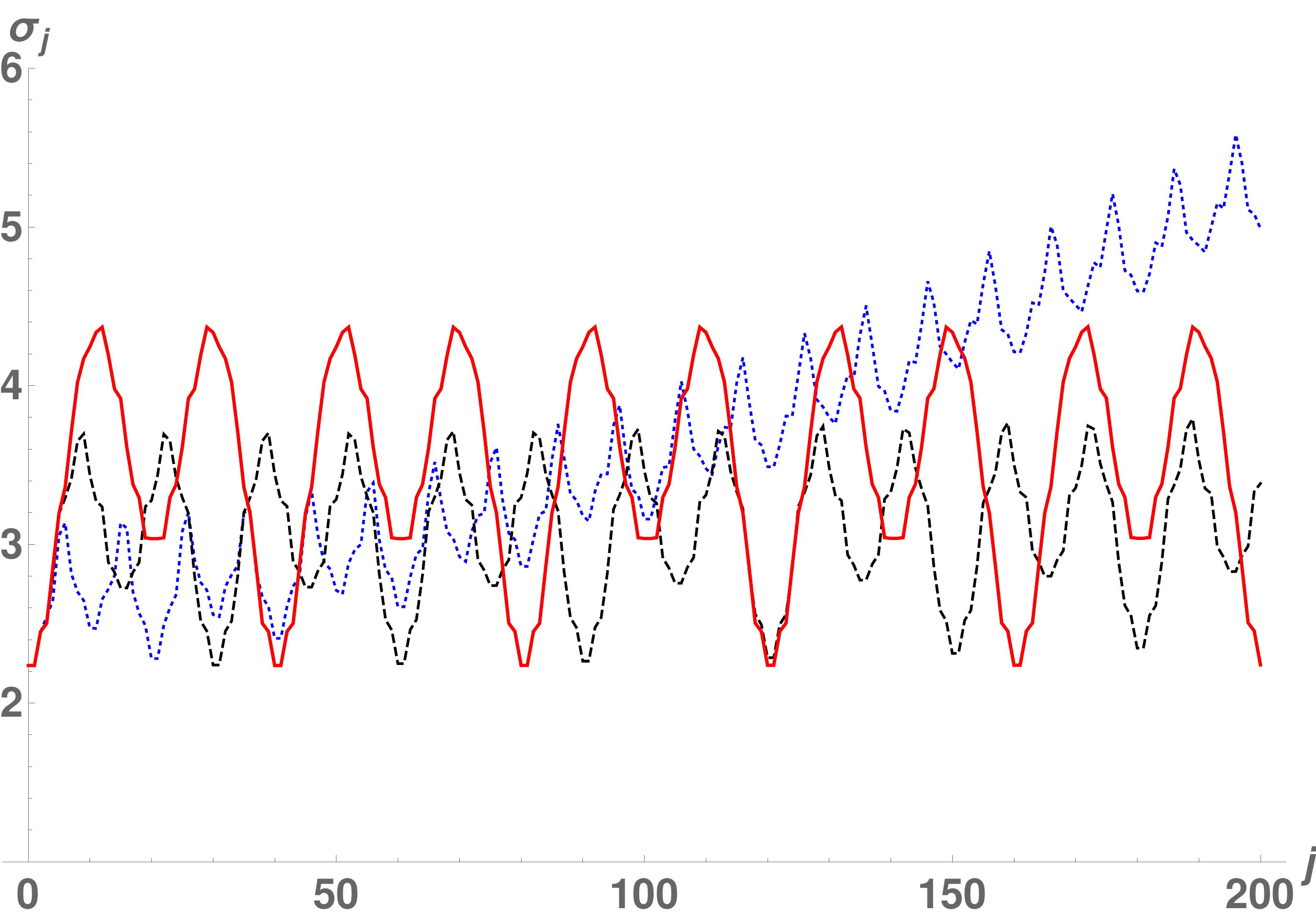}
\par\end{centering}
\caption{Average position $\langle n\rangle_{j}$ (upper panel) and standard
deviation $\sigma_{j}$ (lower panel) for initial Gaussian states,
see Eq.\ \eqref{eq:c_n}, as a function of the time instant $j$,
for different field strenghts: $\phi=2\pi/20$ (blue, dotted line),
$\phi=2\pi/30$ (black, dashed line), and $\phi=2\pi/40$, (red, solid
line). The rest of the initial condition is as in Fig.\ \ref{Fig:lowering_phi}.
\label{Fig:xav203040}}
\end{figure}

The situation described in Sec.\ \ref{subsubsec:breathing} both for TBHs and DQWs, is that observed for an initially localized walker. 
Now, in TBHs, the situation changes when one considers extended initial conditions, since in this case one obtains a wavepacket which oscillates around the starting position, which is the semi-classical prediction.
The question that arises is whether this kind of behavior has a parallelism in DQWs.
To investigate this question, we have first performed numerical simulations considering an initial product state with Gaussian spatial part,
\begin{equation} \label{eq:Initial_state}
\Psi_{0,n} \equiv \escp{x_{n}}{\Psi_0}  = c_{n}\ket s \, ,
\end{equation}
where $\ket s$ is the initial coin state, and where
\begin{equation}  \label{eq:c_n}
c_{n}=\frac{e^{-\beta n^{2}}}{\sqrt{\vartheta_{3}\left(0,e^{-2\beta}\right)}} \, ,
\end{equation}
describes a Gaussian, which, because it is defined on a discrete space, must be normalized, in order to get $\sum_{n}|c_{n}|^{2}=1$, via
\begin{equation}
\vartheta_{3}\left(u,q\right)=1+\sum_{n=1}^{\infty}q^{n^{2}}\cos (2nu) \, ,
\end{equation}
the third Jacobi theta function.  

For "high" values of $\phi$, i.e., of the order of $2\pi$, there
is a rich phenomenology. 
However, as $\phi$ is lowered, the contour plots of the probability
distribution tend to converge towards a simple pattern, namely, oscillations
of a localized particle with the Bloch period $T_{\mathrm{Bloch}}(\phi)$.
This tendency is illustrated in Fig. \ref{Fig:lowering_phi}, where
we present $P_{j,n}$ for a strong and a weak $\phi$. As can be seen
from the plots, the oscillations become smoother as $\phi$ decreases
-- and of course the Bloch period increases. Similar observations
can be extracted from Fig.\ \ref{Fig:xav203040}, by observing the
curves that show the average position $\langle n\rangle_{j}$ which
are obtained by lowering $\phi$. 

An important message is obtained by observing the lower panel in this
Figure, which plots the standard deviation $\sigma_{j}$ as a function
of the timestep $j$ for the same values $\phi=2\pi/20$, $\phi=2\pi/30$,
and $\phi=2\pi/40$. Since $\phi/2\pi$ is a rational number in these
plots, one should observe a long-term ballistic behavior with (imperfect)
revivals, as proven in \cite{ced13} for all kind of initial states
(including the initial Gaussian states considered here). This effect
is apparent for $\phi=2\pi/20$, but becomes less visible for lower
values of $\phi$. 

\begin{figure}
\includegraphics[width=8cm]{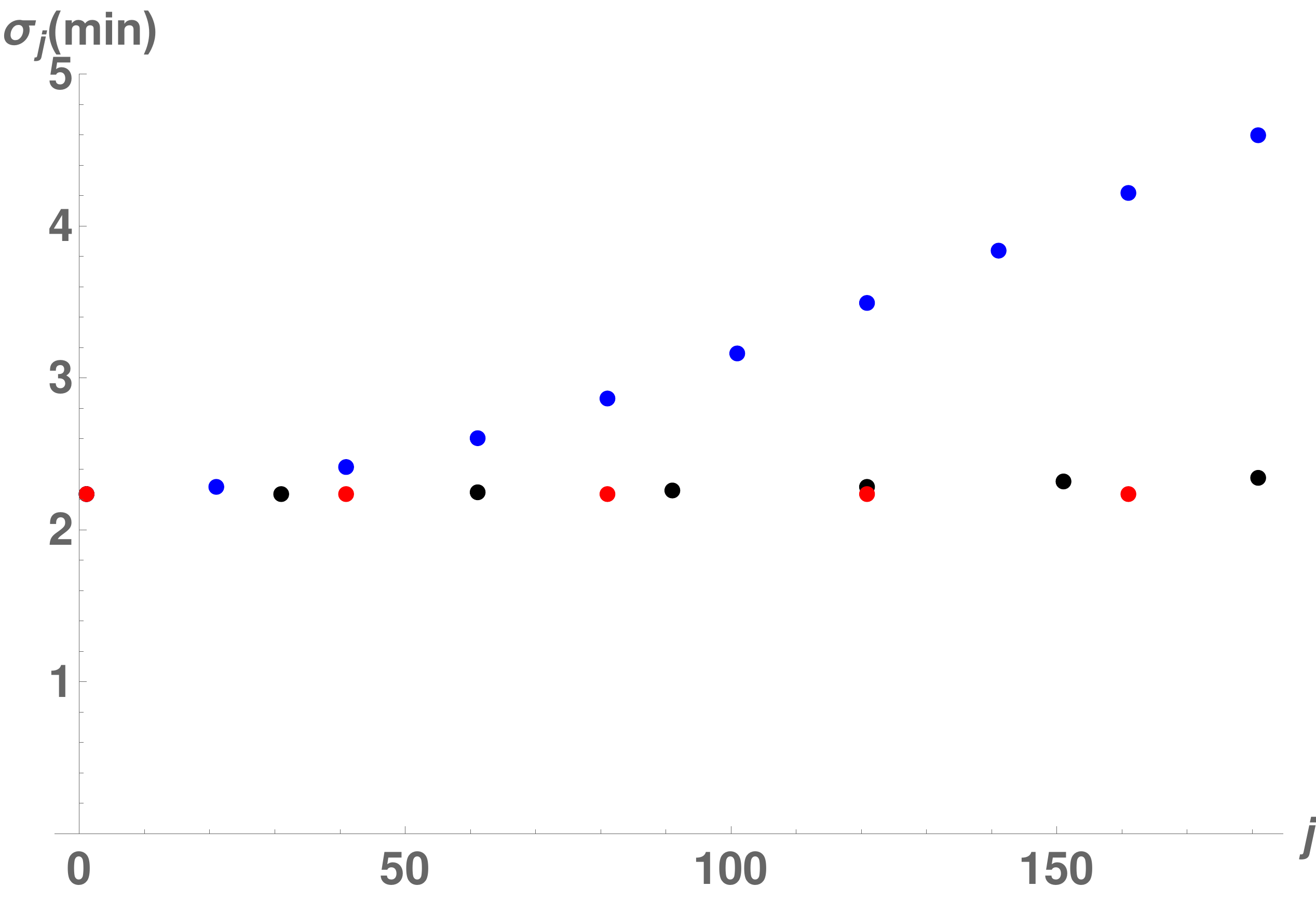}

\caption{The dots represent the values of the standard deviation $\sigma_{j}$
corresponding to the minima in the lower panel of Fig. \ref{Fig:xav203040},
using the same conventions for colors for the different values of
$\phi$.}

\label{Fig:minima_sigmaj}
\end{figure}

Fig. \ref{Fig:minima_sigmaj} plots the minima of $\sigma_{j}$, as
obtained from the curves represented in the lower panel of Fig. \ref{Fig:xav203040}.
As can be seen from this figure, this magnitude increases with $j$
(in this sense, the revivals become more imperfect as the time step
increases). However, this tendency becomes quickly unobservable as
$\phi$ is decreased for the same number of time steps. In fact, for
$\phi=2\pi/60$ (not shown) the minima, during the same total duration,
are constant within the machine precision. 

These observations indicate a \textit{convergence of the dynamics}
for weak fields, and encourages us to derive an analytical expression
in the regime we are interested in: small $\phi$ and wide initial
conditions. This is the goal of the next Section.

\section{\label{sec:Continuous-time-limit} Continuous-time approximation for weak electric
fields and wide initial conditions}

\subsection{Introduction: the two branches arising from the two-step dynamics}

We start by rewriting Eq.\ (\ref{eq:Dynamicsk}) as
\begin{equation}
\tilde{\Psi}_{j+1}(k+\phi)=W_{0}(k)\tilde{\Psi}_j(k) \, .\label{eq:Dynamicskplus}
\end{equation}
Since $W_{0}(k)$ is unitary, one also has
\begin{equation}
\tilde{\Psi}_{j-1}(k-\phi)=W_{0}^{\dagger}(k-\phi)\tilde{\Psi}_j(k) \, ,\label{eq:Dynamicskminus}
\end{equation}
with $W_{0}^{\dagger}(k)$ the Hermitian conjugate of $W_{0}(k)$.
By combining the above two equations, we arrive to 
\begin{equation}
\tilde{\Psi}_{j+1}(k+\phi)-\tilde{\Psi}_{j-1}(k-\phi)=\Delta_{\phi}(k)\tilde{\Psi}_j(k) \, ,\label{eq:Two-stepdynamics}
\end{equation}
where we have defined
\begin{equation} \label{eq:diff}
\Delta_{\phi}(k)\equiv W_{0}(k)-W_{0}^{\dagger}(k-\phi) \, .
\end{equation}

Equation (\ref{eq:Two-stepdynamics}) involves two time steps, from
$j-1$ to $j+1$, so that we will call it \emph{two-step dynamics}, at variance with the original dynamics Eq.\
(\ref{eq:Psievol}), which involves a single time step, so that we will call it \emph{one-step dynamics}.
The two-step dynamics takes as input, not one, but two initial conditions.
It is easy to show by induction that the two-step and the one-step dynamics are equivalent provided that we choose, for the two-step dynamics, the second initial condition given by the one-step dynamics,
\begin{equation} \label{eq:equiv}
\ket{\Psi_{1}} = W_0(\hat{k}) \ket{\Psi_{0}} \, ,
\end{equation}
which we will assume in what follows.

This two-step dynamics has already been considered in Ref.\ \cite{Knight2003}
(see also Ref.\ \cite{Romanelli2004}) for the free DQW, and in Ref.\ \cite{Bauls2006}
for the electric DQW, although only in physical space, and using
the temporal gauge for the electric field \cite{DMD14}.
Let us make use of the findings of these works: 
We assume that one can find two auxiliary partial states $A^{\pm}_j(k)$ such that
\begin{equation}
\tilde{\Psi}_j(k)=A^{+}_j(k)+(-1)^{j}A^{-}_j(k) \, ,\label{eq:PsiwithAplusminus}
\end{equation}
with $A^{\pm}_j(k)$ satisfying
\begin{equation} \label{eq:Two-step-A}
A^{\pm}_{j+1}(k+\phi)-A^{\pm}_{j-1}(k-\phi)=\pm\Delta_{\phi}(k)A^{\pm}_j(k) \, .
\end{equation}

\subsection{Continuous-time approximation for weak electric fields and wide initial conditions}

\subsubsection{Introduction}

Without electric field, it is known since Strauch's work \cite{Strauch06b}
that spacetime-uniform DQWs are well approximated via spacetime-uniform
CQWs -- which, if their graph is a lattice, are nothing but TBHs
--, if the DQW is chosen lazy, i.e., if the coin operator is almost
a coin flip. One can see with analytical arguments or simply with
numerical checks that, if the electric field of the electric DQW is
chosen weak, then the above-mentioned continuous-time approximation
will hold at least for some time and certain initial conditions, and,
this time, the Hamiltonian of the CQW will be an \emph{electric} TBH,
i.e., a standard TBH with an additional superimposed electric field.
Notice that, from its definition, Eq.\ \eqref{eq:DefW}, one can
see a periodicity of $2\pi$ in the variable $\phi$, which can therefore
be restricted to the interval $[-\pi,\pi[$. According to this, a
field will be called "weak" if it satisfies $|\phi|\ll\pi$. Now,
as we have seen in Sec.\ \ref{subsec:Bloch-oscillations}, DQW Bloch
oscillations actually hold even if the DQW is not chosen lazy, i.e.,
for \emph{arbitrary} coin operators, and a continuous-time approximation
is suggested by the smoothness of these oscillations when the initial
condition is wide, see Fig.\ \ref{Fig:lowering_phi}. One can analytically
readily see, without electric field, that a continuous-time approximation
will hold at least for some time, \emph{even if the DQW is not chosen
lazy, provided the initial condition is wide}; We give details on
this in Appendix \ref{app:CTapprox}. Here, we are going to assume
that the continuous-time approximation holds for wide-enough initial
conditions, which will enable us to derive a formula that we will
directly compare to numerical simulations of the original, discrete-time
dynamics.

We introduce a continuous time variable $t$ and a time step $\tau >0$, and assume that $\tilde{\Psi}_j(k)$ coincides with a continuous function of time, $\tilde{\Psi}(\cdot,k) : t \mapsto \tilde{\Psi}(t,k)$, at instants $t_j \equiv j \tau$, i.e., $\tilde{\Psi}_j(k) = \tilde{\Psi}(t,k)$.
We assume $\tilde{\Psi}(t,k)$ to be twice differentiable in $t$.
Taylor expanding now the left-hand side of Eq.\ \eqref{eq:Two-step-A} at first order in both $\tau$ (continuous time) and $\phi$ (weak electric field), we arrive to 
\begin{equation} \label{eq:CT}
i \left( \frac{\tau}{\phi} {\partial_t A^{\pm}(t,k)}+ {\partial_k A^{\pm}(t,k)} \right)=\pm\frac{i}{2\phi}\Delta_{\phi}(k)A^{\pm}(t,k) \, ,
\end{equation}
which is a partial differential equation in both $t$ and $k$, where we have included the imaginary unit for convenience, and where the notations in the assumed continuous-time approximation should be clear from the context.

\subsubsection{Characteristic curve and two-step Hamiltonian}

The left-hand side of Eq.\ \eqref{eq:CT} possesses the characteristic curve $k_{t} \equiv k-\phi t/\tau$ : Indeed, by introducing the new functions
\begin{equation}
\breve{A}^{\pm}(k,k_{t}) \equiv A^{\pm}(t,k) \, ,
\end{equation} one obtains 
\begin{equation}
i {\partial_k \breve{A}^{\pm}(k,k_{t})}=\pm\frac{i}{2\phi}\Delta_{\phi}(k)\breve{A}^{\pm}(k,k_{t}) \, ,\label{eq:DerPsitilde}
\end{equation}
which is an ordinary differential equation in $k$, i.e., time has been removed from the original equation, Eq.\ \eqref{eq:CT}.
Let us examine Eq.\ (\ref{eq:DerPsitilde}) in more detail.
It takes the form of a Schrödinger equation in which $k$ plays the role of time, up to the fact that $\Delta_{\phi}(k)$, see Eq.\ (\ref{eq:diff}), is not Hermitian.
Let us show that this lack of Hermiticity is actually only due to the fact that we have not yet performed the small-field approximation on $\Delta_{\phi}(k)$, which is needed for the consistency of the expansion.
If we expand $\Delta_{\phi}(k)$ in $\phi$ and only keep the lowest contribution in $\phi$,  Eq.\ (\ref{eq:DerPsitilde}) becomes
\begin{equation}
i {\partial_k \breve{A}^{\pm}(k,k_{t})} =\pm\frac{1}{\phi}H_{2}(k)\breve{A}^{\pm}(k,k_{t}) \, , \label{eq:DifH2}
\end{equation}
where we have introduced an operator that we call \emph{two-step Hamiltonian},  
\begin{equation}
H_{2}(k)\equiv\frac{i}{2}\left[W_{0}(k)-W_{0}^{\dagger}(k)\right] \, ,\label{eq:DefH2}
\end{equation}
which is Hermitian.

This two-step Hamiltonian differs from the so-called \emph{effective Hamiltonian} $H_{1}(k)$ that can be defined from $W_{0}(k)$ via
\begin{equation}
W_{0}(k)\equiv e^{-i\tau H_{1}(k)} \, . \label{eq:DefH1}
\end{equation}
Now, we show in Appendix \ref{app:H1H2} that $H_{1}(k)$ and  $H_{2}(k)$ are actually proportional,
\begin{equation}
H_{1}(k)=\frac{\omega(k)}{\tau\sin\omega(k)}H_{2}(k) \, ,\label{eq:H1H2}
\end{equation}
where $e^{\pm i \omega(k)}$ are the eigenvalues of the matrix $W_0(k)$, assumed in $\mathrm{SU}(2)$.
Eq.\ (\ref{eq:H1H2}) is a useful result for the free DQW, since the direct calculation of $H_{1}(k)$ via Eq.\ \eqref{eq:DefH1}  is quite involved (see, e.g., Mallick's work \cite{Mallick2019}), while the calculation of $H_{2}(k)$ defined by Eq.\ \eqref{eq:DefH2} is a trivial task.
%

\subsubsection{Explicit solutions in quasimomentum space, for an arbitrary coin operator}

The solution of Eq.\ (\ref{eq:DifH2}) can be written, formally, as
\begin{equation}
A^{\pm}(t,k) \equiv \breve{A}^{\pm}(k,k_{t}) = V_{\phi}^{\pm}(k)S^{\pm}(k_{t}) \, ,\label{eq:PsiVandS}
\end{equation}
where $S^{\pm}(k)$ is a state that can be obtained from the initial conditions (see below), and where we have defined the unitary operators
\begin{equation} \label{eq:op}
V_{\phi}^{\pm}(k)=\mathcal{T}\exp \! \! \left[\mp\frac{i}{\phi}\int_{0}^{k}dpH_{2}(p)\right] \, .
\end{equation}
We have to introduce the $p$-ordering $\mathcal{T}$ because in general $H_{2}(p)$ and $H_{2}(p')$ do not commute for  $p\neq p'$.
Equation \eqref{eq:op} determines, together with the knowledge of $A^{\pm}(0,k)$, the state $S^{\pm}(k)$ as
\begin{equation} \label{eq:condini}
S^{\pm}(k)\equiv [V_{\phi}^{\pm}(k)]^{\dagger}A^{\pm}(0,k) \, .
\end{equation}
Putting everything together, i.e., plugging Eq.\ \eqref{eq:condini} into Eq.\ \eqref{eq:PsiVandS}, and the resulting expression in Eq.\ \eqref{eq:PsiwithAplusminus}, we arrive to
\begin{equation}
\tilde{\Psi}(t,k)  =  \sum_{\alpha=+,-} \alpha^{t/\tau}V_{\phi}^{\alpha}(k)[V_{\phi}^{\alpha}(k_{t})]^{\dagger}A^{\alpha}(0,k_{t}) \, .\label{eq:PsiwithAAdagger}
\end{equation}

Let us now show that we can express the initial condition $A^{\pm}(0,k)$ via $\tilde{\Psi}(0,k)$.
Equation \eqref{eq:PsiwithAplusminus} taken for $j=0$ and $j=1$ yields, respectively,
\begin{align}
\tilde{\Psi}(0,k) &= A^+(0,k) + A^-(0,k) \\
\tilde{\Psi}(\tau,k) &= A^+(\tau,k) - A^-(\tau,k)  \\
&= A^+(0,k) - A^-(0,k) + O(\tau) \, , \label{eq:2}
\end{align}
so that, dropping the $O(\tau)$, i.e., at lowest order in $\tau$ (continuous-time approximation), and
recalling that $\tilde{\Psi}(\tau,k) = W_{0}(k)\tilde{\Psi}(0,k)$ (equivalence condition between the one-step dynamics and the two-step one, see around Eq.\ \eqref{eq:equiv}), we obtain
\begin{equation}
A^{\pm}(0,k) =  \frac{1}{2} \left[ 1 \pm W_{0}(k) \right] \tilde{\Psi}(0,k) \, . \label{eq:At0withW0}
\end{equation}
Recall that if we insist on the possibility to have arbitrary angles $\theta$, the continuous-time approximation only holds in the long-wavelength approximation, and this comes from the structure of mere free walk, as explained in Appendix \ref{app:CTapprox} and as already noticed in Ref.\ \cite{Knight2003}.
Choosing, furthermore, a weak electric field, is necessary for this approximation to hold in the present case.

\subsubsection{Choice of a wide, i.e., spatially extended initial condition, and associated simplifications}

In what follows, we will consider a product initial state, with the external-degree-of-freedom part defined in Eq.\ (\ref{eq:Initial_state}).
In quasimomentum space, it translates into
\begin{equation} \label{eq:Psi(k,0)}
\tilde{\Psi}(0,k)=g(k)\ket s \, ,
\end{equation}
where 
\begin{equation}
g(k)=\sum_{n} c_{n} e^{-ikn} =\sqrt{\frac{\pi}{\beta}}\, e^{-\frac{k^{2}}{4\beta}} \, \frac{\vartheta_{3} \! \! \left(\frac{ik\pi}{2\beta},e^{-\frac{\pi^{2}}{\beta}}\right)}{\vartheta_{3}\! \left(0,e^{-2\beta}\right)} \, ,\label{eq:g(k)}
\end{equation}
 is a quasimomentum amplitude distribution, which verifies $\int_{-\pi}^{\pi}\frac{dk}{2\pi}|g(k)|^{2}=1$. 

Assuming $g(k)$ peaked around $k=0$, i.e., the initial distribution to be spatially extended, allows for a further simplification in Eq.\ (\ref{eq:At0withW0}), namely,
\begin{equation}
A^{\pm}(0,k) =  g(k)\Lambda^{\pm}\ket s,
\end{equation}
where
\begin{equation}
\Lambda^{\pm} \equiv \frac{1}{2}[1 \pm W_{0}(0)]=\frac{1}{2}[1 \pm C] \, ,
\end{equation}
are the projectors on the eigenspaces of $W_{0}(0)=C$, and thus as projectors verify $\Lambda^{+}\Lambda^{+}=\Lambda^{+}$, $\Lambda^{-}\Lambda^{-}=\Lambda^{-}$, $\Lambda^{+}\Lambda^{-}=\Lambda^{-}\Lambda^{+}=0$, and $\Lambda^{+}+\Lambda^{-}=1$.
Using the above expressions, one can recast Eq.\ (\ref{eq:PsiwithAAdagger}) as 
\begin{equation}
\tilde{\Psi}(t,k)  =  U(t,k)\tilde{\Psi}(0,k_t) \, ,\label{eq:PsiwithUkt}
\end{equation}
having defined
\begin{equation}
U(t,k)  \equiv  \sum_{\alpha=+,-}\alpha^{t/\tau}V_{\phi}^{\alpha}(k)[V_{\phi}^{\alpha}(k_{t})]^{\dagger}\Lambda^{\alpha} \, .\label{eq:U(k,t)}
\end{equation}
One can easily check that $U(t,k)$ is a unitary operator, and this holds because the $\Lambda^{\pm}$ are projectors, which is guaranteed if the initial condition is wide.

\subsubsection{Connection with a certain tight-binding Hamiltonian}

With our choice for the coin operator, Eq.\ (\ref{eq:ChoiceC}),  one obtains 
\begin{equation}
H_{2}(k)=\cos\theta\sin k\,\, \mathbf{1}_2,\label{eq:H2TBM}
\end{equation}
with $\mathbf{1}_2$ the $2\times2$ identity matrix.
This result allows to establish a remarkably simple connection with TBHs.

Indeed, consider the following lattice Hamiltonian,
\begin{equation}
H(\hat{k})\equiv iJ(\hat{T}-\hat{T}^{\dagger}) \, ,\label{eq:HTB}
\end{equation}
where $\hat{T} \equiv e^{-{i}\hat{k}} = \sum_{n} \ket{n+1} \! \! \bra n$ is the hopping operator to the right, $J \in \mathbb{R}$ is
a constant, and recall that we have taken the lattice spacing $a=1$.
In quasimomentum space, one obtains 
\begin{equation}
H(k)=2J\sin k,\label{eq:HTBk}
\end{equation}
which corresponds to the factor multiplying $\mathbf{1}_2$ in Eq.\ (\ref{eq:H2TBM}),
provided we make the following correspondence,
\begin{equation} \label{eq:correspondence}
\cos\theta \, \leftrightarrow \, 2J \, .
\end{equation} 

\subsubsection{Explicit solutions in quasimomentum space}

Since $H_{2}(k)$ in Eq.\ (\ref{eq:H2TBM}) is proportional to $\mathbf{1}_2$, we can omit this trivial factor in what follows, and then immediately get
\begin{equation}
V_{\phi}^{\pm}(k)=\exp \! \! \left[\mp\frac{i}{\phi}f(k) \right],\label{eq:V_phi_scalar}
\end{equation}
where
\begin{equation}
f(k)\equiv\cos\theta (1-\cos k) \, .
\end{equation}
 We also have
\begin{equation}
\Lambda^{\pm}=\left(\begin{array}{cc}
\cos^{2}\frac{\theta}{2} & \pm\frac{1}{2}\sin\theta\\
\pm\frac{1}{2}\sin\theta & \sin^{2}\frac{\theta}{2}
\end{array}\right).
\end{equation}
Making use of Eqs.\ (\ref{eq:U(k,t)}), (\ref{eq:Psi(k,0)}), and (\ref{eq:V_phi_scalar}),
we can finally write 
\begin{equation} \label{eq:Psi(k,t)withF} 
\tilde{\Psi}(t,k)=\big[F^{+}(t,k) \, \Lambda^{+}+(-1)^{t}F^{-}(t,k) \, \Lambda^{-} \big] \ket s \, ,
\end{equation}
where
\begin{equation} \label{eq:FPlusminusk}
F^{\pm}(t,k)\equiv g(k_{t})e^{\mp\frac{i}{\phi}[f(k)-f(k_{t})]} \, ,
\end{equation}
and where we have taken $\tau=1$ as the unit time step.
Using some trigonometric identities one can write 
\begin{equation}
f(k)-f(k_{t})=-2\cos(\theta) \sin(k-\phi t/2)\sin(\phi t/2) \, .
\end{equation} 
%

\subsubsection{Explicit solutions in position space}

We know the following relationship,
\begin{equation}
\Psi_n(t)\equiv\escp{x_{n}}{\Psi(t)}=\frac{1}{2\pi}\int_{-\pi}^{\pi}dk \, e^{ikn}\tilde{\Psi}(t,k) \, ,
\end{equation}
so that taking the inverse Fourier transform of Eq.\ \eqref{eq:Psi(k,t)withF} yields
\begin{equation} \label{eq:Psi(n,t)_aprox}
\Psi_n(t)=\big[F^{+}_n(t)\, \Lambda^{+}+(-1)^{t}F^{-}_n(t) \, \Lambda^{-}\big]\ket s \, ,
\end{equation}
where 
\begin{equation}
F^{\pm}_n(t)\equiv\frac{1}{2\pi}\int_{-\pi}^{\pi} dk \, e^{ikn} F^{\pm}(t,k) \, .
\end{equation}

Making use of Eq.\ (\ref{eq:g(k)}) we arrive, after some tedious but straightforward algebra, to the final expressions
{\small
\begin{align}
F^{+}_n(t)&=\sum_{l}c_{l} \, e^{i(n+l)\phi t/2} \, J_{n-l}\big(2 \gamma \sin \tfrac{\phi t}{2} \big)  \\
F^{-}_n(t)&=\sum_{l} \,c_{l} \, e^{i(n+l) \phi t/2} \,J_{-(n-l)}\big(2 \gamma \sin \tfrac{\phi t}{2} \big) \, ,
\end{align}}
with the $c_l$s the initial coefficients of the ``wavefunction'', see Eq.\ \eqref{eq:Initial_state}, having introduced
\begin{equation}
\gamma \equiv \frac{\cos \theta}{\phi} \, ,
\end{equation}
and where we have made use of the property $(-1)^p J_p(X) = J_{-p}(X)$ holding for $p \in \mathbb{Z}$, the $J_p$s being Bessel functions of the first kind \cite{Gradshteyn:1702455}.
Notice that these expressions with Bessel functions are \emph{exactly} those that appear in TBH models, if we make the correspondance $\cos\theta\leftrightarrow 2J$.
More precisely, the probability density reads
\begin{align} \label{eq:P(n,t)_aprox}
P_n(t) &\equiv  \Psi^{\dag}_n(t) \Psi_n(t) \nonumber \\
&=  |F^{+}_n(t)|^2 \, \langle s | \Lambda^{+} |  s \rangle + | F^{-}_n(t)|^2 \, \langle s | \Lambda^{-} |  s \rangle   \, , 
\end{align}
so that one clearly sees that the solution is simply the probability sum of two counter-propagating solutions of the previously mentioned TBH (see Eq.\ (18) in Ref.\ \cite{Hartmann2004}), weighted by the probabilities of going in one or the other direction given the initial coin state $\ket s$.
Notice that the fact that there are no interference terms between the two branches in the probability density is because the $\Lambda^{\pm}$ are projectors.

\subsubsection{First comparison between the DQW and its continuous-time approximation}

\begin{figure}
\begin{centering}
\includegraphics[width=7cm]{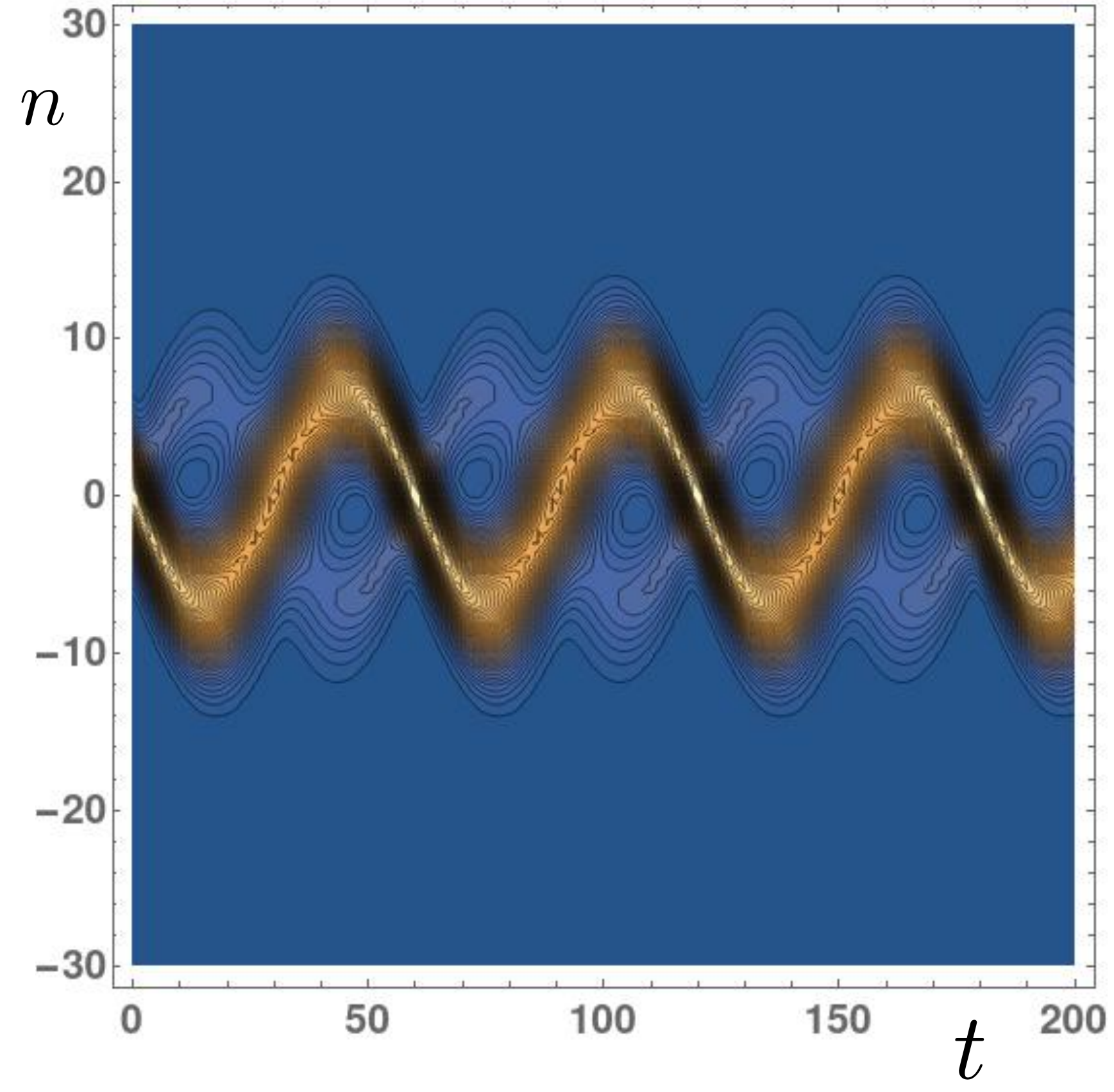}
\par\end{centering}
\caption{Contour plots corresponding to the probability density $P_n(t)$,
as a function of the lattice site $n$ and the time instant $t$, as
obtained from Eq.\ (\ref{eq:Psi(n,t)_aprox}). The initial conditions
are the same as in the lower panel of Fig. \ref{Fig:lowering_phi}. \label{Fig:Gaussian2piover60_aprox} }
\end{figure}

\begin{figure}
\begin{centering}
\includegraphics[width=6cm]{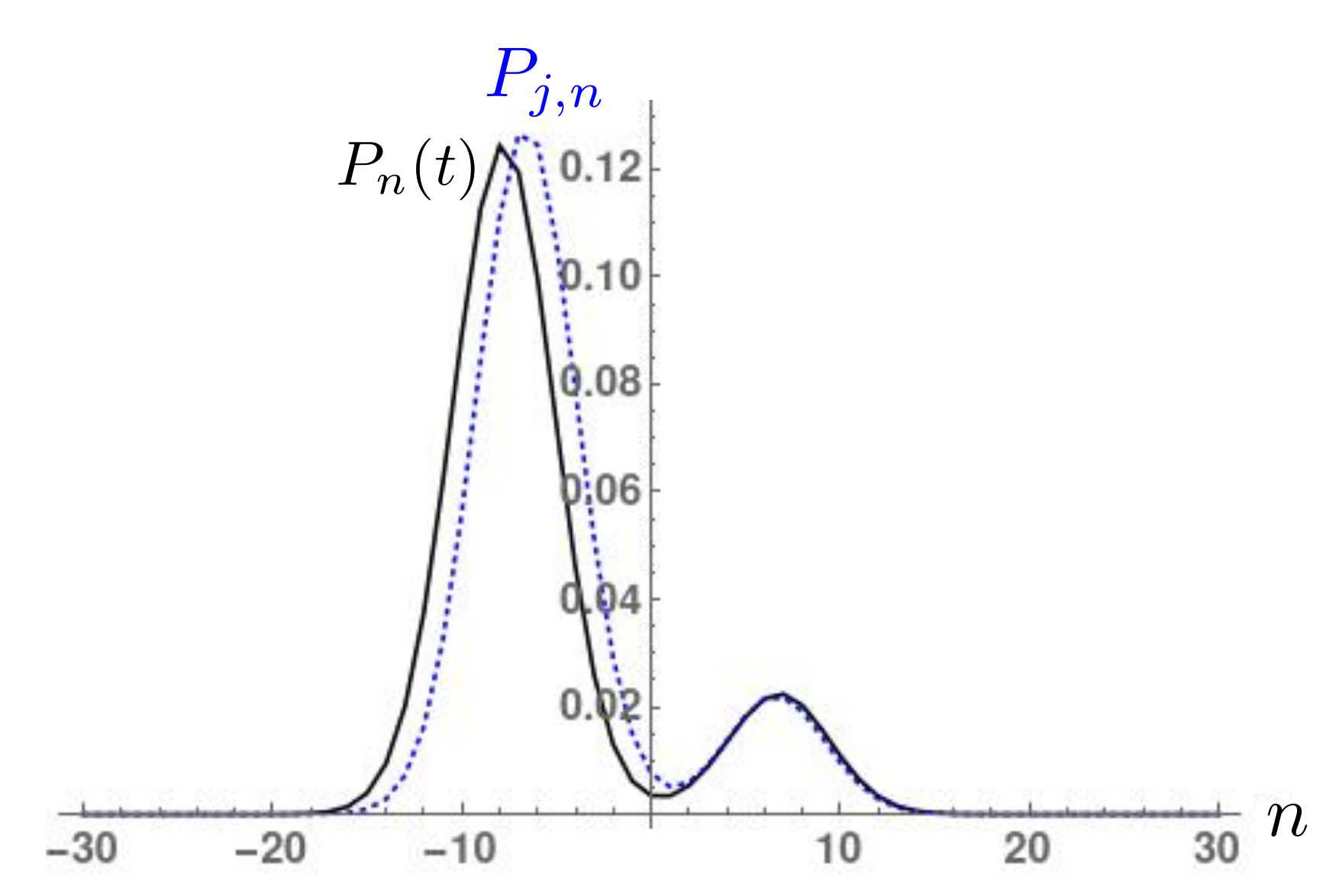} 
\par\end{centering}
\begin{centering}
\includegraphics[width=6cm]{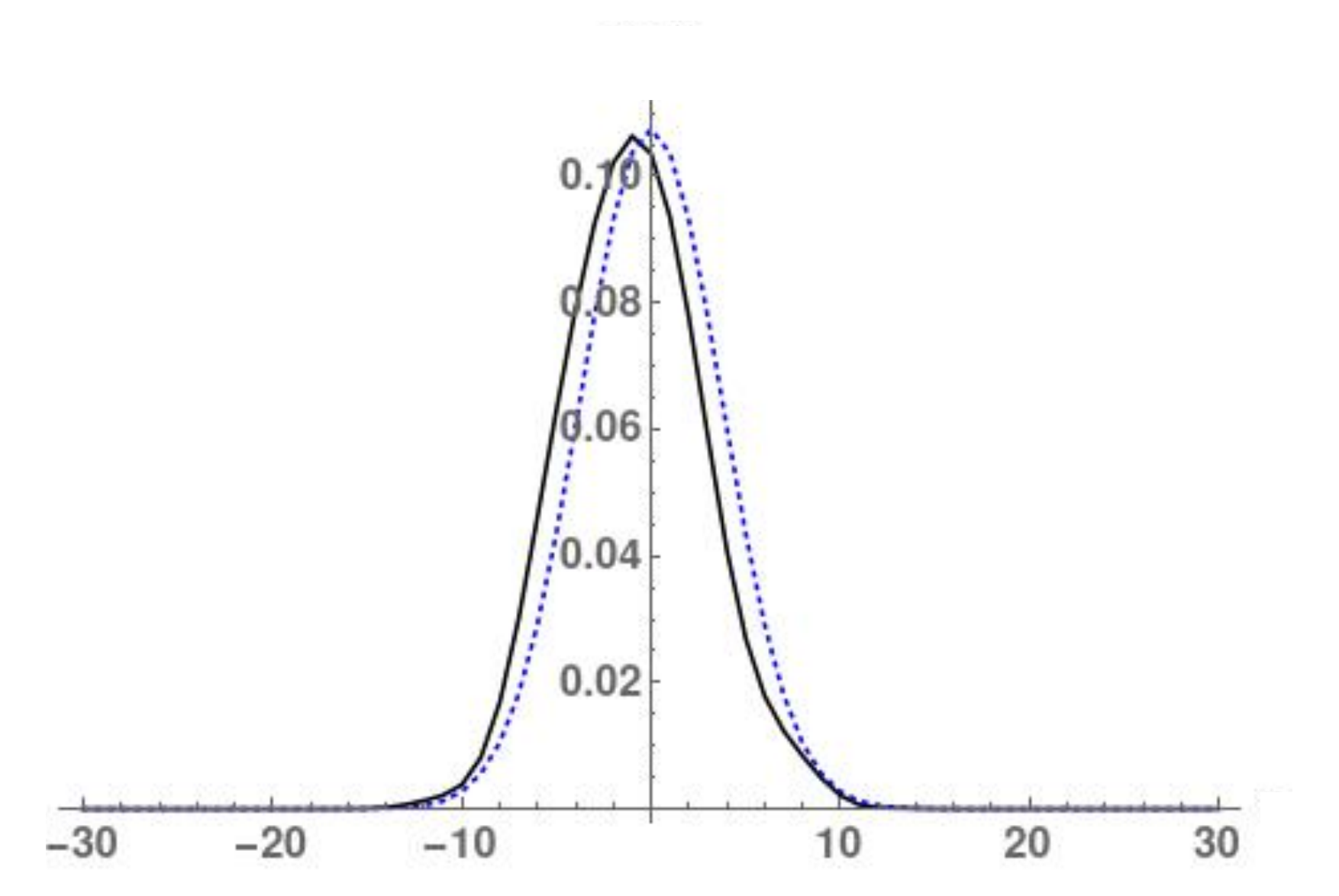}
\par\end{centering}
\begin{centering}
\includegraphics[width=6cm]{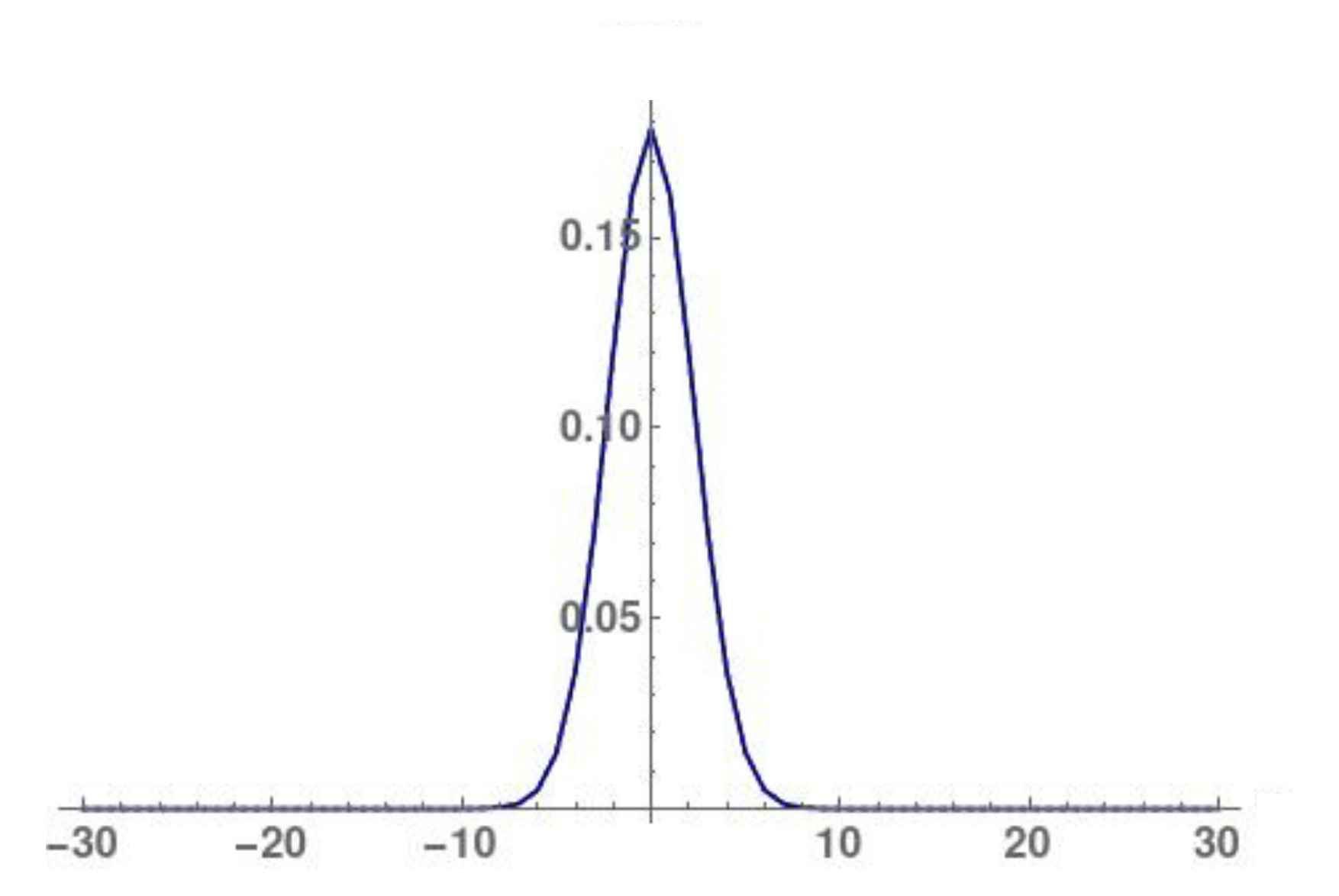}
\par\end{centering}
\caption{Different snapshots of the probability $P_n(t)$, as obtained from Eq.\ (\ref{eq:Psi(n,t)_aprox}) (black solid line), with the same parameters as in Fig.\ \ref{Fig:Gaussian2piover60_aprox}, in comparison with the exact simulation that gives $P_{j,n}$ (blue dotted line). The different panels correspond, from top to bottom, to $t=15$, $t=30$ and $t=60$. In this and similar plots, the points corresponding to different values of $n$ have been joined for a better visualization. \label{Fig:Snapshots}}
\end{figure}

Fig.\ \ref{Fig:Gaussian2piover60_aprox} is a contour plot obtained with the same conditions as in Fig.\ \ref{Fig:lowering_phi}, but using the approximated analytical result, Eq.\ (\ref{eq:Psi(n,t)_aprox}).
As can be immediately seen, both plots agree visually very well.

In order to evaluate with more accuracy the agreement observed in comparing both contour plots, we show in Fig.\ (\ref{Fig:Snapshots}) three different snapshots of the probability $P_n(t)$, as obtained from our approximated expression, Eq.\ (\ref{eq:Psi(n,t)_aprox}), in comparison with the exact numerical simulation.
As can be appreciated from this figure, the degree of agreement can be different, depending on the exact time step, although it is generally good.

This figure also shows that the initial Gaussian exhibits a second small peak.
The presence of this secondary peak arises from the combination of the two terms in Eq.\ (\ref{eq:Psi(n,t)_aprox}), and is also visible both in Figs.\ \ref{Fig:lowering_phi} and \ref{Fig:Gaussian2piover60_aprox}.
In fact, by choosing the initial coin state $\ket s$ to be one of the eigenstates of the coin operator $C$, then either $\Lambda^{+}\ket s=0$ or $\Lambda^{-}\ket s=0$, so that the probability would not show such a secondary peak. 

Another important result that can be obtained from the above approximated expressions is an analytical formula for both $\langle n\rangle_{t}$ and $\langle n^{2}\rangle_{t}$.
Both calculations are performed in Appendix \ref{app:Average-position-and}, and show and oscillatory result, with a period dictated by $T_{\mathrm{Bloch}}(\phi)$.
As a consequence, our formulae predict that the DQW remains localized for any \emph{weak enough} value of $\phi$ and a wide initial condition. 

We would like to conclude this section with an important observation.
We notice that some expressions like the operator $V_{\phi}^{\pm}(k)$ show a singular behavior at $\phi=0$ (see Eq.\ (\ref{eq:V_phi_scalar})).
Such a singular limit already appears in tight-binding models with an external electric field, where the Wannier-Stark eigenstates present this singularity \cite{Hartmann2004}.
However, observable magnitudes as, for example, $\langle n\rangle_{t}$ or $\langle n^{2}\rangle_{t}$, do possess a well defined limit as $\phi\rightarrow0$, which can be obtained from Eqs.\ (\ref{eq:n(t)final}) and (\ref{eq:n2(t)final}), giving (i)
\begin{equation}
\langle n\rangle_{t}\underset{\phi\rightarrow 0}{\longrightarrow} \kappa \, e^{-\beta/2}t\sin\theta \, ,
\end{equation}
which depends on the initial coin state via $\kappa$ defined in Eq.\ \eqref{eq:defd}, and (ii)
\begin{equation}
\langle n^{2}\rangle_{t}\underset{\phi\rightarrow 0}{\longrightarrow}\frac{1}{4\beta}+t^{2}\cos^{2} \! \theta \, ,
\end{equation}
which does not depend on the initial coin state.
These two formula are consistent with the ballistic propagation observed in the free DQW.

\section{Measure of probability agreement} \label{sec:Measure-of-probability}

\begin{figure}
\begin{centering}
\includegraphics[width=7cm]{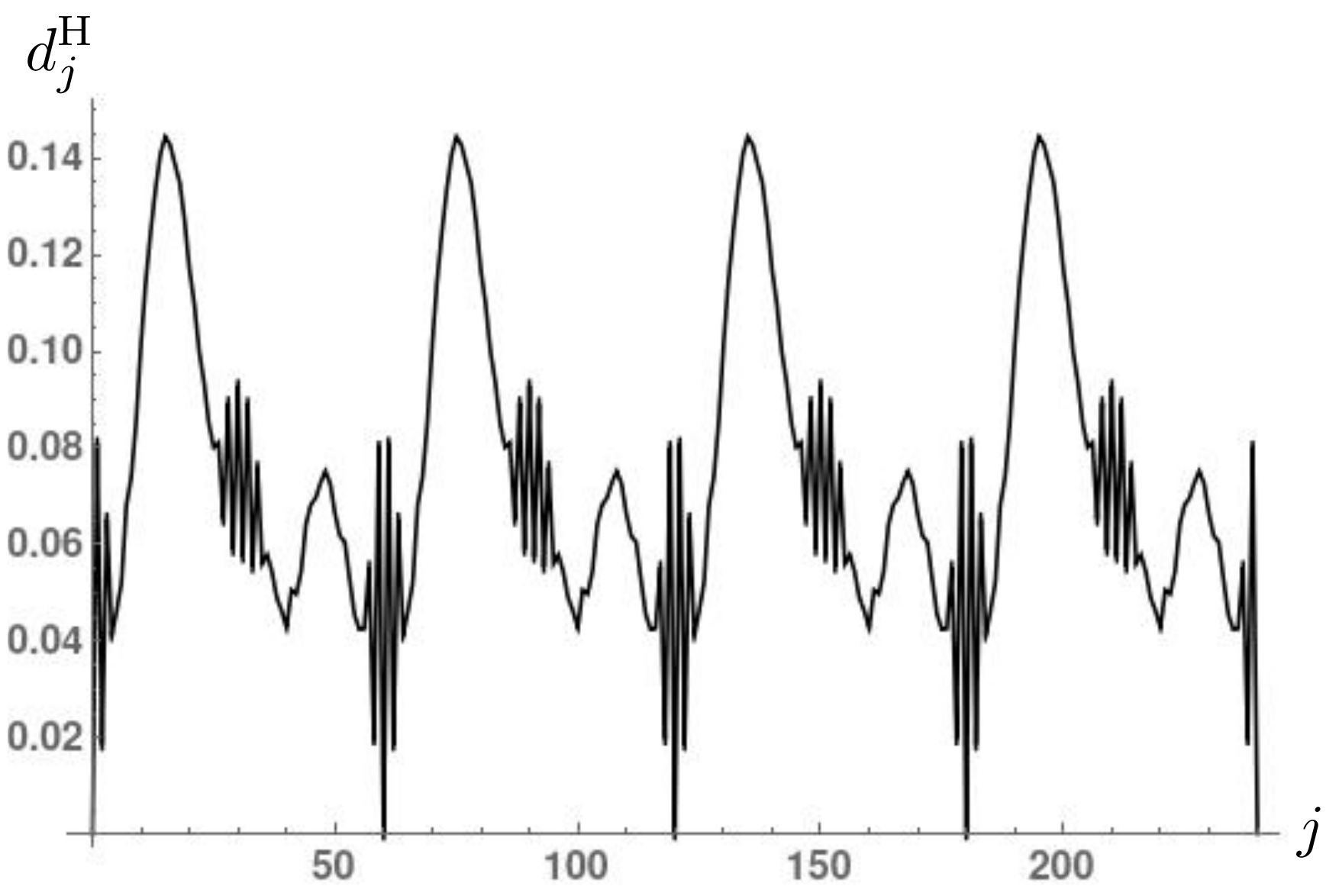}
\includegraphics[width=7cm]{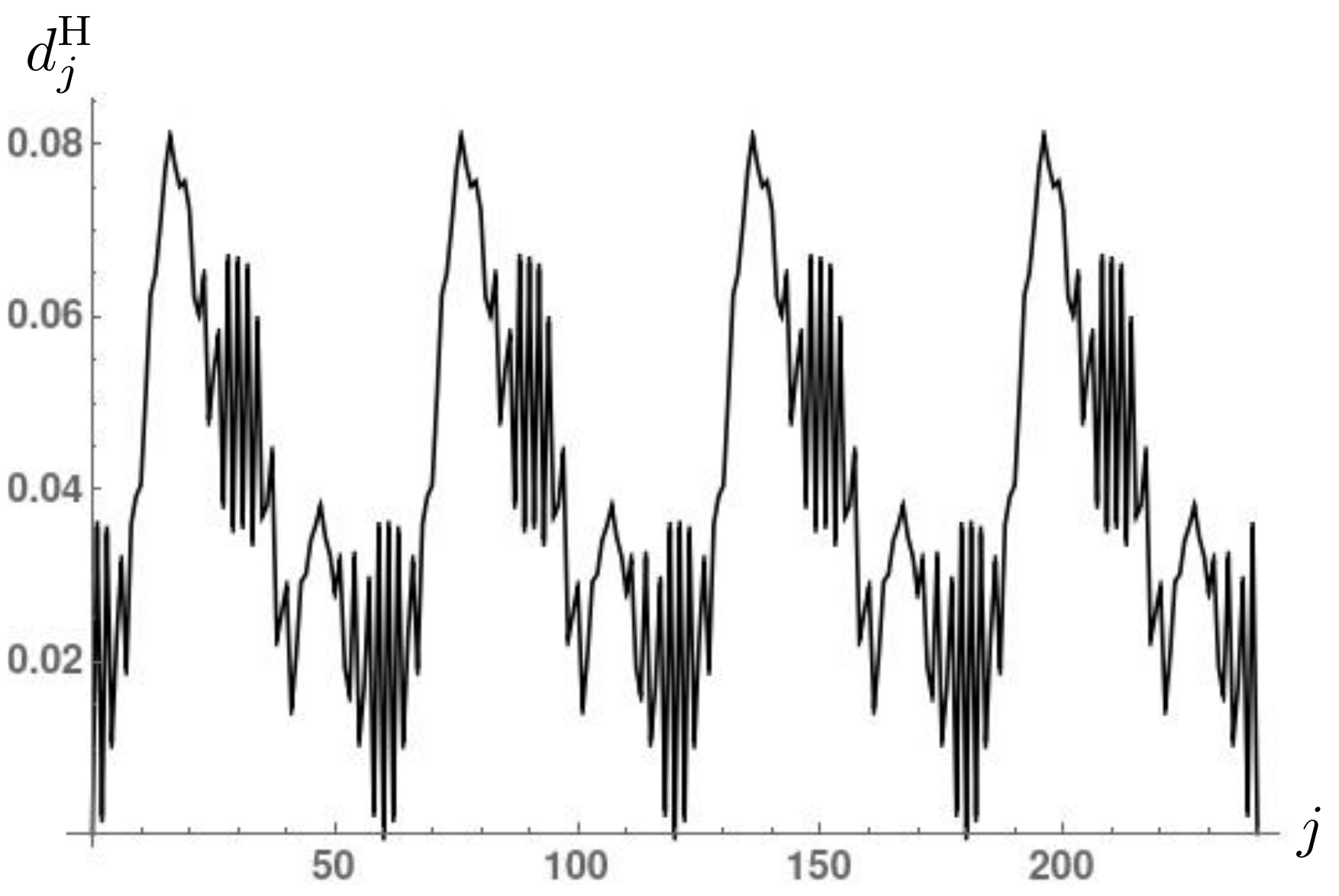}
\par\end{centering}
\caption{Hellinger distance $d^{\mathrm{H}}_j$ between the exact and the approximated probability
distributions, as a function of the time instant $j \in \mathbb{N}$. We have used the
same parameters as in Fig.\ \ref{Fig:Gaussian2piover60_aprox} for the top plot, i.e., in particular, $\beta=0.05$, and also for the bottom plot except for the choice $\beta=0.01$. \label{Fig:He1}}
\end{figure}

As we have observed in the previous section, we have obtained a good agreement between the calculated probability distribution $P_n(t)$, see Eq.\ \eqref{eq:P(n,t)_aprox}, that comes from the approximated result of Eq.\ (\ref{eq:Psi(n,t)_aprox}), and the exact simulation.
In this section, we would like to quantify this degree of agreement and, more importantly, to study how it changes with time.
This question becomes very important if one wants to extract some conclusions about the long-term behavior of the DQW in the considered regime (weak electric field and wide initial condition). 

In order to compare both probability distributions, we need some distance measure.
There are, in fact, several measures available which are specially suited to compare two probability distributions, such as
the Hellinger distance, the total variation distance and the Kolmogorov-Smirnov distance (see \citet{Dobson2004} for a definition of these terms).
All of them give extremely similar plots, so we have concentrated on the Hellinger distance, which we write, for a given time instant $j$,
as 
\begin{equation}
d^{\mathrm{H}}_j=\sqrt{1-\sum_{n}\sqrt{P^{\mathrm{ex}}_{j,n} P^{\mathrm{apr}}_n(t_j)}} \, ,
\end{equation}
where $P^{\mathrm{ex}}_{j,n}$ is the exact probability distribution, whereas $P^{\mathrm{apr}}_n(t_j)$ is obtained from Eq.\ (\ref{eq:Psi(n,t)_aprox}) and evaluated at time $t_j = j \tau$, with $\tau = 1$.
This distance will be bounded as $0\leq d^{\mathrm{H}}_j \leq1 \,\,\,\forall t$.

In the top plot of Fig.\ \ref{Fig:He1} is shown $d_{j}^{\mathrm{H}}$
for the same conditions used in previous figures. Several comments
can be made about this figure. Within a numerical precision of the
order $10^{-8}$, it is a periodic function of period $T_{\mathrm{Bloch}}(\phi)$,
as expected from the results observed in the previous section. It
reaches a minimum value (numerically compatible with zero) at values
of the time instant which are multiples of $T_{\mathrm{Bloch}}(\phi)$
(compare with the last panel in Fig. \ref{Fig:Snapshots}). The maxima
in $d_{j}^{\mathrm{H}}$ appear at time instants $j\simeq T_{\mathrm{Bloch}}(\phi)/4$
(modulo $T_{\mathrm{Bloch}}(\phi)$), with a constant value (considering
the announced precision). It is important to mention that changing
$\phi$ to an arbitrary close value does not change significantly
the above plot, except for the fact that multiples or submultiples
of $T_{\mathrm{Bloch}}(\phi)$ may not correspond to integer time
instants $j$, a fact that introduces some small changes in the appearance
of the figure. For this reason, we restrict ourselves to choices of
$\phi$ giving an integer $T_{\mathrm{Bloch}}(\phi)$. Another important
point to be made, is that the above observations have been extracted
from calculations which involve only a small number of Bloch periods.
As discussed in Sect. II, as the value of $\phi$ is lowered, one
quickly approaches a regime where the qualitative features of the
probability distribution do not depend on $\phi$, and one has to
consider much longer times (which would require a large amount of
computational resources) to observe the differences. 

We have explored different values of the field and of the initial conditions.
Changing the intensity $\phi$ does not imply a significant modification of the top plot of Fig.\ \ref{Fig:He1} (of course, the period will be changed).
A similar statement can be said about a modification of the initial coin state.
Obviously, these statements only refer to the observed distance between both probability distributions, not to the overall evolution of the probability, which can be tailored by modifying the initial parameters, as discussed in Sec. \ref{sec:Continuous-time-limit}.
The width of the initial Gaussian, however, has an important impact on the above figure.
This can be appreciated on the bottom plot of Fig.\ \ref{Fig:He1}, which has been obtained with $\beta=0.01$, which implies a wider initial Gaussian.
As it can be seen, this introduces a better agreement between the exact and the approximated distributions.

\section{Conclusions} \label{sec:Conclusions}

In this work is presented a semi-analytical study of the well-known
electric DQW on the line \cite{ced13}, Eq. \eqref{eq:DefW}, in the
case where (i) the electric field $\phi$ is small: $|\phi|\ll\pi$,
and (ii) the initial condition is spatially extended. It is shown
that, in such a regime, the dynamics of the electric DQW corresponds
to semi-classical oscillations which are well approximated by an analytical,
continuous-time formula that is the sum of two counter-propagating
solutions of a certain electric TBH, that is, well-known semi-classical
Bloch oscillations of a localized particle \cite{Hartmann2004,Dominguez-Adame2010}.
The hopping amplitude of this TBH is the cosine $\cos\theta$ of the
\emph{arbitrary} coin-operator mixing angle $\theta$. The result
is semi-analytical in the sense that the quality of the analytical
approximation is evaluated numerically, via an appropiated distance
measure. The price to pay for the arbitrariness of $\theta$ is that
the initial condition must be wide. If one wishes the continuous-time
approximation to hold for spatially localized initial conditions,
one needs at least the DQW to be lazy, as suggested by numerical simulations
and by the fact that this has been analytically proven in the case
of a vanishing electric field \cite{Strauch06b}.

\begin{acknowledgments}
We acknowledge illuminating discussions with Christopher Cedzich and
Eugenio Roldán. This work has been funded by the Spanish Ministerio
de Economía, Industria y Competitividad , MINECO-FEDER project FPA2017-84543-P,
SEV-2014-0398 and Generalitat Valenciana grant PROMETEU/2019/087.
We acknowledge support from CSIC Research Platform PTI-001. 
\end{acknowledgments}

\appendix

\section{Simple conditions to approximate a DQW by a CQW} \label{app:CTapprox}

We will derive these conditions for the free walk, $W_0(\hat{k})$, defined in Eq.\ \eqref{eq:free_walk}, which we write $\mathcal{W}$ in the position representation, for the coin operator, defined in Eq.\ \eqref{eq:ChoiceC}.
We have, in the position representation
\begin{subequations}
\begin{align}
\Psi_{j+1,n} &= (\mathcal{W} \Psi_j)_n \\
\Psi_{j-1,n} &= (\mathcal{W}^{\dag} \Psi_j)_n \, ,
\end{align}
\end{subequations}
which yields,
{\small
\begin{align} \label{eq:first}
\Psi_{j+1,p} - \Psi_{j-1,p}& =
 \left(\mathcal{S} \left( \begin{matrix} 
c \, \psi^R_{j} + s  \, \psi^L_{j} \\
s  \, \psi^R_{j} - c  \, \psi^L_{j} \end{matrix}\right) \right)_{\! \! n} - C 
\left( \begin{matrix} 
\psi^R_{j,n+1} \\
\psi^L_{j,n-1} \end{matrix} \right) \, ,
\end{align}}
with the coin components as defined in Eq.\ \eqref{eq:components}, where $\mathcal{S}$ is the position representation of $S(\hat{k})$, and where we have used the notations
\begin{subequations}
\begin{align}
\cos \theta &= c \\
\sin \theta &= s \, .
\end{align}
\end{subequations}
Equation \eqref{eq:first} further results in
{\small
\begin{align}
\Psi_{j+1,n} - \Psi_{j-1,n} = \left( \begin{matrix}
c \, \psi^R_{j,n-1} + s \, \psi^L_{j,n-1} - c \, \psi^R_{j,n+1} - s \, \psi^L_{j,n-1} \\
s \, \psi^R_{j,n+1} - c \, \psi^L_{j,n+1} - s \, \psi^R_{j,n+1} + c \, \psi^L_{j,n-1} 
\end{matrix} \right) \, ,
\end{align}}
which simplifies into the two following (decoupled\footnote{The decoupling arises here because the coin operator $C$ is Hermitian.}) equations
{\small
\begin{subequations}
\begin{align}
\psi^R_{j+1,n} - \psi^R_{j-1,n} &= \, \, \, 
c \, \,  \big( \psi^R_{j,n-1} -  \psi^R_{j,n+1} \big)  \\
\psi^L_{j+1,n} - \psi^L_{j-1,n} &=
- c \big( \psi^L_{j,n+1}  - \, \psi^L_{j,n-1} \big) \, .
\end{align}
\end{subequations}}
We want this discrete-time dynamics to be approximable by a continuous-time one, i.e., we want, for $u=L,R$,
\begin{equation} \label{eq:approx}
| \psi^u_{j+1,n} - \psi^u_{j-1,n} | = |c| | \psi^u_{j,n-1} -  \psi^u_{j,n+1} | \ll | \psi^u_{j-1,n} | \, .
\end{equation}
As Eq.\ \eqref{eq:approx} suggests, a first possibility for this to be satisfied at least for some time, even if the initial condition is not wide in space, is, as shown in Ref.\ \cite{Strauch06b} (we will not give additional information here but refer the reader to that reference), to choose $c$ small enough, i.e., $C$ to be almost a coin flip: If one starts with a Dirac delta at $n=0$ initially, one ends up after two steps with almost the same Dirac delta at $n=0$, the amplitude being diminished by a small amount only, so that the continuous-time approximation is good.
As Eq.\ \eqref{eq:approx} suggests, a second possibility for the continuous-time approximation to hold at least for some time whatever angle $\theta$ we choose, is to take an initial condition which is wide in space, as already considered for the free walk in Ref.\ \cite{Knight2003}, and for the electric walk in the temporal gauge in Ref.\ \cite{Bauls2006} (in the present work, we chose the spatial gauge because it is simpler in quasimomentum space).
Notice that quality of the continuous-time approximation depends on \emph{joint effect} of these \emph{two} conditions; It is only in the two respective limits ($\theta \rightarrow \pi/2$, or infinite-wavelength limit) that one can forget about the other condition.

\section{Relationship between the two-step Hamiltonian $H_2(k)$ and the effective Hamiltonian $H_1(k)$} \label{app:H1H2} 

In this Appendix, we prove Eq.\ (\ref{eq:H1H2}), holding when the free walk $W_{0}(k)$ is special unitary, i.e., belonging to $\mathrm{SU}(2)$, so that its eigenvalues can be written $e^{\pm i\omega(k)}$.
Notice that this condition of special unitarity is not satisfied with the coin operator of Eq.\ (\ref{eq:ChoiceC}).
One can write
\begin{equation}
W_{0}(k)=\mathcal{U}(k)\left(\begin{array}{cc}
e^{-i\omega(k)} & 0\\
0 & e^{i\omega(k)}
\end{array}\right)\mathcal{U}^{\dagger}(k),
\end{equation}
where $\mathcal{U}(k)$ is a unitary matrix containing the eigenvectors of $W_{0}(k)$.

Now, on the one hand, from the definition of $H_1(k)$ in Eq.\ (\ref{eq:DefH1}), and assuming the principal branch for the logarithm, one arrives to
\begin{equation}
H_{1}(k)=\frac{\omega(k)}{\tau}\mathcal{U}(k)\sigma_{z}\mathcal{U}^{\dagger}(k) \, .
\end{equation}
On the other hand, from Eq.\ (\ref{eq:DefH2}) one easily obtains
\begin{equation}
H_{2}(k)=\sin\omega(k)\mathcal{U}(k)\sigma_{z}\mathcal{U}^{\dagger}(k) \, .
\end{equation}
By combining both expressions, Eq.\ (\ref{eq:H1H2}) immediately follows.

\section{Average position and standard deviation} \label{app:Average-position-and}

The average position defined in Eq.\ (\ref{eq:n(t)}) can be more easily calculated in quasimomentum space as
\begin{equation}
\langle n\rangle_{t}=\frac{1}{2\pi}\int_{-\pi}^{\pi}dk \, \tilde{\Psi}^{\dagger}(t,k) \, i \partial_k \tilde{\Psi}(t,k) \, ,
\end{equation}
where $\tilde{\Psi}^{\dagger}(t,k)$ is obtained by transposition and complex conjugation from $\tilde{\Psi}(t,k)$.
Using the expression of $\tilde{\Psi}(t,k)$ given by Eq.\ (\ref{eq:Psi(k,t)withF}), and making use of the properties of the projectors $\Lambda^{\pm}$, one can write
{\small
\begin{align} \label{eq:n(t)withFs}
\langle n\rangle_{t} & =  \frac{i}{2\pi}\int_{-\pi}^{\pi} dk \Big[{F^{+}}^{\ast} \! (t,k) \, \partial_k F^{+}(t,k)  \, \langle s | \Lambda^{+} | s \rangle  \\
 & \ \ \ \ \ \ \ \ \ \ \ \ \ \ \  + {F^{-}}^{\ast} \! (t,k) \,  \partial_k F^{-}(t,k)  \, \langle s | \Lambda^{+} | s \rangle \Big]  \, .  \nonumber 
\end{align}}
Writing the coin state as
\begin{equation}
\ket s \equiv \left( \begin{matrix}
a \\
b
\end{matrix} \right) \, ,
\end{equation}
with the condition that $|a|^{2}+|b|^{2}=1$, Eq.\ (\ref{eq:n(t)withFs}) yields
\begin{align} \label{eq:n(t)tointegrate}
\langle n\rangle_{t}  & =  \frac{1}{2\pi\phi}\int_{-\pi}^{\pi}dk \, g(k_{t}) \Big[ \phi g'(k_{t}) \nonumber \\
 & -  2 i \kappa  \cos(\theta) g(k_{t}) \sin\!\left(\tfrac{\phi t}{2}\right)\cos\!\left(k-\tfrac{\phi t}{2}\right)\Big] \, ,
\end{align}
where
\begin{align} \label{eq:defd}
\kappa \equiv \langle s | C | s \rangle = b^{*}(a\sin\theta-b\cos\theta)+a^{*}(a\cos\theta+b\sin\theta) \, ,
\end{align}
is the mean value of the coin operator $C$ in the initial state $\ket{s}$, which is here a real number because $C$ is Hermitian.
The exact computation of this integral yields an unpractical result because the expression of $g(k)$, see Eq.\ \eqref{eq:g(k)}, is too complicated.
However, since we only consider the long-wavelength limit, we can replace the summation in the definition of $g(k)$ by an integral, which amounts to regard this function as the Fourier transform of the continuous function $c(n)\equiv c_{n}$, where $c_{n}$ is defined by Eq.\ \eqref{eq:c_n}.
In this way, we approximate
\begin{equation}
g(k)\simeq(2\pi\beta)^{1/4}e^{-\frac{k^{2}}{4\beta}} \, .
\end{equation}
Since we are interested in low values of $\beta$, the integration limits in Eq.\ (\ref{eq:n(t)tointegrate}) become $\pm\infty$.
With these approximations, the above integral can be performed, resulting in
\begin{equation}
\langle n\rangle_{t}= \kappa \, e^{-\beta/2}\sin(\theta) \frac{\sin(\phi t)}{\phi} \, . \label{eq:n(t)final}
\end{equation}

A similar procedure can be followed to obtain $\langle n^{2}\rangle_{t}$.
Similarly to Eq.\ (\ref{eq:n(t)withFs}), one has
{\small
\begin{align} \label{eq:n2(t)withFs}
\langle n^{2}\rangle_{t} & =  -\frac{1}{2\pi}\int_{-\pi}^{\pi}dk \Big[{F^{+}}^{*} \! (t,k) \, {\partial^{2}_k}F^{+}(t,k) \,  \langle  s |  \Lambda^{+} | s \rangle  \\
 & \ \ \ \ \ \ \ \ \ \  \ \ \ \ \ \ \  \  + {F^{-}}^{*}\! (t,k) \, \partial^{2}_k F^{-}(t,k)  \, \langle s | \Lambda^{-} | s \rangle \Big] \, . \nonumber
\end{align}}
By performing the same approximations as done in the calculation of $\langle n\rangle_{t}$, we finally obtain the expression
\begin{equation} \label{eq:n2}
\langle n^{2}\rangle_{t}=\frac{1}{4\beta}+2 \cos^{2}(\theta)\frac{\sin^{2} \! \left( \tfrac{\phi t}{2}  \right)}{\phi^2} \left(1+e^{-2\beta}\cos(\phi t)\right) \, .
\end{equation}
Using the fact that the initial Gaussian, see Eq.\ \eqref{eq:c_n}, is wide, i.e., that $\beta$ is small, so that $e^{-2\beta} \simeq 1$, the previous result, Eq.\ \eqref{eq:n2}, can be further approximated by
\begin{equation}
\langle n^{2}\rangle_{t}\simeq\frac{1}{4\beta}+\cos^{2}(\theta)\frac{\sin^{2}(\phi t)}{\phi^2} \, . \label{eq:n2(t)final}
\end{equation}

\bibliographystyle{apsrev4-1}
\input{Paper_Bloch_v4.bbl}

\end{document}

%% file: Paper_Bloch_v4.bbl
%